# Low-energy, ultrafast spin reorientation at competing hybrid interfaces with tunable operating temperature

*Servet Ozdemir\*, Matthew Rogers, Jaka Strohsack, Hari Babu Vasili, Manuel Valvidares, Thahabh Haddadi, Parvathy Harikumar, David O'Regan, Gilberto Teobaldi, Timothy Moorsom, Mannan Ali, Gavin Burnell, B J Hickey, Tomaz Mertelj, Oscar Cespedes\**

S. Ozdemir, M. Rogers, H. B. Vasili, T. Haddadi, M. Ali, G. Burnell, B. J. Hickey, O. Cespedes

School of Physics and Astronomy, University of Leeds, Leeds, UK

E-mail: s.ozdemir@leeds.ac.uk, o.cespedes@leeds.ac.uk

J. Strohsack, T. Mertelj

Department of Complex Matter, Jozef Stefan Institute and Center of Excellence on Nanoscience and Nanotechnology Nanocenter (CENN Nanocenter), Jamova 39, Ljubljana, Slovenia

M. Valvidares

ALBA Synchrotron Light Source, E-08290 Cerdanyol del Valles, Barcelona, Spain

P. Harikumar, D. O'Regan

School of Physics and CRANN Institute, Trinity College Dublin, Dublin, Ireland

G. Teobaldi

Scientific Computing Department, STFC UKRI, Rutherford Appleton Laboratory, Didcot, UK

T. Moorsom

School of Chemical and Process Engineering, University of Leeds, Leeds, UK

(Dated: 27 May 2025)

**Information can be stored in magnetic materials by encoding with the direction of the magnetic moment of elements. A figure of merit for these systems is the energy needed to change the information – rewrite the storage by changing the magnetic moment. Organic molecules offer a playground to manipulate spin order, with metallo-molecular interfaces being a promising direction for sustainable devices. Here, we demonstrate a spin reorientation transition in molecular interfaces of high magnetisation 3d ferromagnetic films due to a competition between a perpendicular magnetic anisotropy (PMA) induced by a heavy metal that dominates at high temperatures, and an in-plane anisotropy generated by molecular coupling at low temperatures. The transition can be tuned around room temperature by varying the ferromagnet thickness (1.4 to 1.9 nm) or the choice of molecular overlayer, with the organic molecules being $C_{60}$, hydrogen and metal (Cu, Co) phthalocyanines. Near the transition temperature, the magnetisation easy axis can be switched with a small energy input, either electrically with a current density of $10^5 \, \text{A/cm}^2$, or optically by a fs laser pulse of fluence as low as 0.12 mJ/cm², suggesting heat**

**assisted technology applications. Magnetic dichroism measurements point toward a phase transition at the organic interface being responsible for the spin reorientation transition.**

## 1. Introduction

Conventional spin reorientation transitions take place when the easy axis of magnetization rotates with respect to the crystallographic axes due to changes in temperature, pressure or magnetic field. The effect has been widely studied since the 1940s[1] for both rare earth systems[2] and magnetic semiconductors[3]. The exponential increase of digital content generated and exchanged as we approach the yottabyte era, combined with environmental objectives, have seen information technologies focus on memory and storage devices made with sustainable materials that allow increased density, lower power consumption and reduced cooling requirements[4]. The search is hampered by the relatively large energies and long times needed to either rotate the magnetisation in conventional hard disks and STT/SOT memories[5,6], or to erase it in heat-assisted storage devices[7]. Hence, the incorporation of new mechanisms such as spin-reorientation transitions into these technologies is an attractive possibility to reduce power consumption. Dilute magnetic semiconductors have been suggested as contenders where a spin reorientation transition takes place, but low operational temperatures and small magnetisation hamper their application[3,8]. Switching that depends on crystallographic axes requires high crystal quality that is difficult to reproduce in commercial nanodevices. Therefore, finding a system where the magnetisation easy axis can be switched within a narrow, controlled temperature range and independently of the crystal structure would be highly beneficial. Interfacial spin dynamics of common transition metals with cheap, eco-friendly organic molecules give rise to emergent functionalities ideal for the observation of novel magnetic phenomena such as a large interfacial tunnel magnetoresistance[9], ferromagnetism onset[10], as well as magnetic hardening[11,12]. Here, we report on metallo-molecular interfaces with an easy axis switching, or spin reorientation transition[1,2] around room temperature with low-energy electrical or optical inputs that lead to high magnetisation changes in ps timescales. The spin reorientation transition temperature, Ts, is controlled through the ferromagnet thickness as well as the molecule used, enabling tunability of the operational temperature as well as device compatibility with standard room temperature information storage and computing technologies.

## 2. Results and Discussion

### 2.1. Universal magnetic hardening and spin-reorientation transition at magneto-molecular interfaces

A comparative study involving molecules of different geometric and magnetic properties is needed to understand the magnetic interactions at the interface[13–15], see Figure 1a-c.  Although different crystal structures are possible, Phthalocyanine (Pc) molecules have been found to grow as α-phase in a flat lying orientation at the interface on various substrates[16], particularly on (111) metal surfaces[17–19]. $C_{60}$ and $H_2Pc$ are diamagnetic, whereas metal phthalocyanines (CuPc, CoPc) have shown magnetic order at low temperatures[20–22]. We find that, despite their varying intrinsic properties, Pt(111)/Co(t≤ 2 nm)/molecular systems exhibit an in-plane magnetisation accompanied by magnetic hardening and very large coercivities of the order of 1 T at low temperatures (Figure 1d). By contrast, reference Pt(111)/Co(t ≤ 2 nm) samples without competing molecular interfaces show

PMA with an out-of-plane (OOP) easy axis at all temperatures (Figure S1, Supporting Information)[23]. The large coercivity is thought to be due to hybridisation and spin polarised charge transfer to the molecular interface (also known as a 'spinterface'[24] ) which induces pinning[15] of the magnetic moment of Co. After the first magnetisation reversal at low temperatures, the coercivity of all samples with a molecular interface reduces to 0.5-0.6 T, which is still one to two orders of magnitude larger than in plain Co films, confirming magnetic hardening of the hybridised Co layer[12]. All four molecular interfaces share this behaviour, suggesting that the origin of the magnetic hardening (the large coercivity) is the same for all samples. Additionally, the influence of central metal ions of CoPc and CuPc molecules can be ruled out to be at the origin of the hardening, because $H_2Pc$ and $C_{60}$ molecules do not have these central metal ions. Equally the spherical geometry of $C_{60}$[15] can be ruled out because phthalocyanine molecules are flat. There have been other cross-molecule studies at Co interfaces where identical spin-polarisation effects at the interface have been observed[25]. The common feature between all molecules used in this study is the hexagonal-pentagonal carbon unit. Density functional theory (DFT) calculations for $Co/C_{60}$ interfaces have shown $C_{60}$ molecules to hybridise with Co atoms in an orientation involving both hexagonal-pentagonal faces[12,15]. Furthermore, studies on $H_2Pc$ metal interfaces have shown the spin polarisation effects focussed on hexagonal-pentagonal units[26]. Remarkably, a simple geometrical comparison of fullerenes and Pcs leads one to find that 4 hexagon-pentagon units are present in the same area of approximately ≈1.54 $nm^2$ for both, enabling d-π orbital hybridisation with a similar number of Co atoms. A spin polarisation of $0.1 - 1$ $\mu_B$ per hexagon-pentagon unit has been observed on fullerenes grown on ferromagnetic metal substrates[12].

The structural and electronic properties of $H_2Pc$ molecules and their interaction with the Co surface were evaluated using density functional theory (DFT) calculations. In the optimized structure, the Co atoms beneath the Pc molecule were found to show slight displacements (Figure S2, Supporting Information), indicating a perturbation due to chemisorption. The PDOS (projected density of states) peaks of Co d-orbitals and C-p orbitals were found to overlap substantially in the same energy range (Figure S3, Supporting Information) indicating hybridization between the metal and the molecule at the hexagonal-pentagonal unit. The strong in-plane anisotropy in Pt/Co/molecular samples at 10 K is clear from changes in the remanence and coercivity in measured hysteresis loops at different angles of magnetic field with respect to the sample plane (Figures S4, Supporting Information). The radical modifications in the magnetic properties of the Co layer are attributed to changes to the valence state of Co close to the surface due to charge transfer and metal-molecule orbital hybridisation (see Table S1, Supporting Information).

Further to the low temperature molecular hybridisation induced hardening, a spin-reorientation transition at higher temperatures is evidenced by the in-plane remanence curves shown in Figure 1e. The onset temperature in 1.7 nm thick Co films ranges from (200±5) K to (285±5) K depending on the molecular overlayer. Although the central-ions are not part of the hexagon-pentagon units that hybridise with the metal to lead to observed effects, we believe they may play a role in influencing the degree of hybridisation of the π-orbitals in the hexagonal-pentagonal units, hence leading to the observed ranges of temperatures for planar magnetisation onset. A perpendicular easy axis at T=400 K is confirmed in OOP hysteresis loops for all samples (Figure 1f). Accompanied by the onset of the PMA, there is an enhancement in the OOP coercivity for all organic molecules when compared to the reference sample. This agrees with previous room temperature measurements in $C_{60}$[27], although the magnitude of the effect is molecule specific.

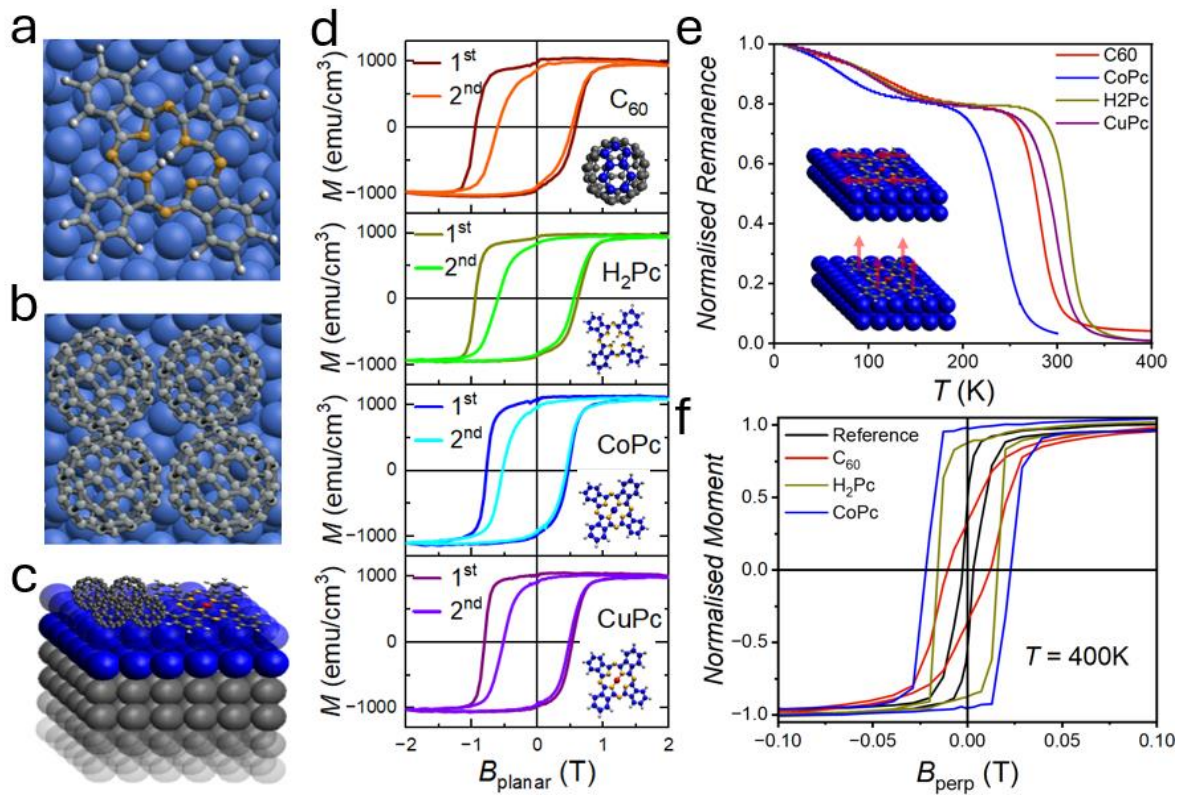

**Figure 1.** Magnetic hardening and spin orientation transition at metallo-molecular thin film interfaces. **a,** A $H_2$-Pc molecule (diameter ~ 1.4 nm) lying in a planar orientation on a Co surface. **b,** Four $C_{60}$ molecules (diameter ~0.7 nm) hybridised with Co surface on hexagon-pentagon faces occupying the same area as a single Pc molecule. **c,** Thin film structure depicted with the Pt seed layer (grey), Co-layer (blue) with metal Pc and $C_{60}$ molecules on the Co surface. **d,** In-plane magnetisation loops ($1^{st}$ and $2^{nd}$ loops depicted in different colours) of ≈1.7 nm Co thickness Pt/Co/$C_{60}$, Pt/Co/$H_2$Pc, Pt/Co/CoPc and Pt/Co/CuPc film structures measured at T=10 K post field cooling at 2 T (corresponding molecules to each curve shown on the insets). **e,** Normalised in-plane magnetic remanence measured post field cooling at 2 T for films of $C_{60}$, $H_2$Pc, CoPc and CuPc interfaced with Pt/Co(1.7 nm) structure with schematics on the inset depicting the spin reorientation transition. **f,** Normalised perpendicular to plane hysteresis loops measured at T=400 K for reference Pt/Co(1.7 nm) films as well as Pt/Co/$C_{60}$, $H_2$Pc and CoPc interfaces showing an enhancement in coercivity.

## 2.2. Cobalt thickness and electrical current induced tunability of spin-reorientation transition

The tunability of the spin reorientation transition was studied for Co films with thicknesses ranging between 1.4 and 1.9 nm and interfaced with CuPc. We find that the switching temperature increases with increased Co film thickness, Figure 2a. For Co films of less than 1.4 nm thickness, OOP magnetisation was exhibited at all measurable temperatures up to 400 K. This indicates that the spin orbit mediated magnetic coupling of the bottom Pt interface is stronger but has a faster thickness decay in comparison with the top molecular interface effect.

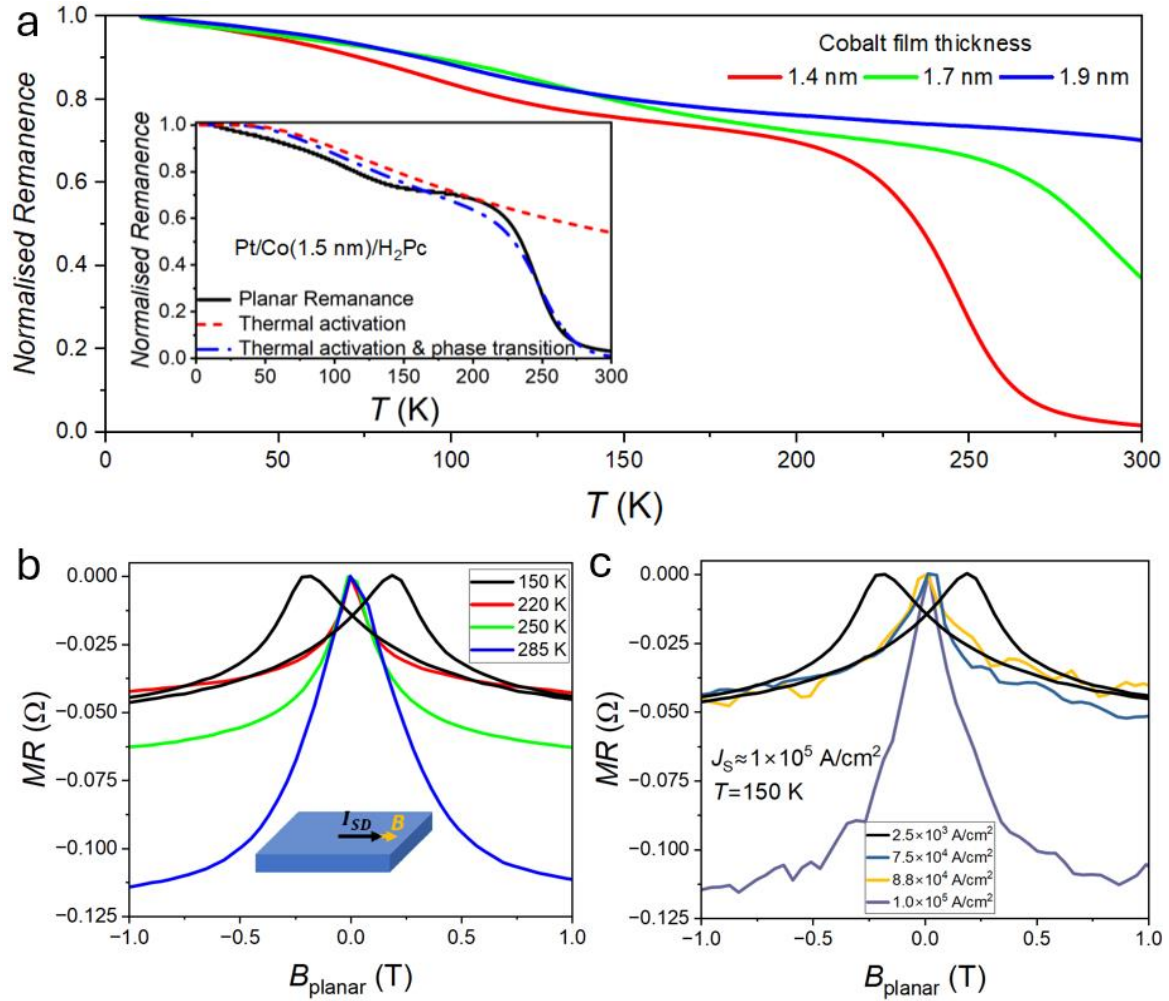

**Figure 2. Cobalt thickness dependence of the spin reorientation transition and its electrical current control. a**, Planar magnetic remanence curves showing Co thickness dependent spin-reorientation transition onset at Pt/Co/CuPc metallo-molecular structures, with the inset of phenomenological fits to a remanence curve for a Pt/Co/H₂Pc structure. **b**, Temperature dependence of magnetoresistance on a Pt/Co(1.7 nm)/CoPc structure measured with a current density of $2.5 \times 10^3 \, A/cm^2$ during Anisotropic Magnetoresistance (AMR) field sweeps, with the field applied parallel to the electrical current (the inset schematic), showing greater negative magnetoresistance as temperature is increased and spin reorientation transition takes place. **c**, AMR field sweeps measured at varying current densities on the Pt/Co(1.7 nm)/CoPc structure at $T$=150 K suggesting an induced spin reorientation transition at a switching current density of $\approx 1 \times 10^5 \, A/cm^2$.

The in-plane easy axis onset can be approximated by a combination of thermally activated remanence with a logistic function (Supporting Information, Note 1). The fit yields an activation energy on the order of 18 meV for a Pt/Co(1.5 nm)/H₂Pc(15 nm) interface (the inset in Figure 2a). Switching temperatures, defined in the phenomenological model as the point at which the IP magnetisation is 50%, measured in different samples were found to be as low as (245±5) K, with an upper limit above 300 K (Figure 1e and 2a, Supporting Information, Table S2). The Pt(111)/Co interface anisotropy constant in the absence of molecules was estimated to be $\cong 0.8 \, mJ/m^2$ below $T_S$ (Figure S6, Supporting Information). This energy needed to switch the anisotropy axis is promising for low-power operation in scalable devices. The thermally activated remanence at $T<T_S$ suggests that the magnetisation axis is settled, with a competition between the temperature-dependent in-plane interfacial molecular anisotropy mediated by hexagon-pentagon π-3d hybridisation, and the

5p-3d hybridised Pt/Co interface. The spin orbit coupling induced perpendicular anisotropy generated at the bottom surface dominates at T>$T_S$ due to molecular vibrations, rotation and repositioning reducing the metallo-molecular coupling.

The narrow temperature window across which the transition occurs, and its tunability around 300 K depending on the Co-layer thickness, points us towards electrical control of the spin-reorientation. To test this, we carried out temperature dependent Anisotropic Magneto-Resistance (AMR) measurements near $T_S$ in a Pt/Co(1.7nm)/CoPc sample (Figure 2b). Such AMR measurements, where an in-plane magnetic field is rotated from parallel (0°) to perpendicular to the current (90°) configuration (Figure S7, Supporting Information) have been utilised to monitor spin-reorientation transitions in smaller samples of exfoliated systems[28]. Below $T_S$, the in-plane coercivity at 0° is observed in the butterfly-like AMR curve. Above $T_S$, there is an increase in the negative AMR as the magnetisation easy axis switches to out of plane. This enables the AMR measurements to act as a probe of the easy axis direction (Figure S8a, Supporting Information, shows a comparison of AMR vs IP magnetisation remanence). A series of high current magnetic field sweeps were carried out at 150 K, well below the $T_S$ of 245 K for this heterostructure as shown in Figure 2c. The easy axis switching from planar to OOP direction was observed at a current density of $J_S = 1 \times 10^5$ A/cm$^2$. We note the abrupt jump in negative magnetoresistance from -0.046 Ω to -0.114 Ω within a current window of $\pm 1.2 \times 10^4$ A/cm$^2$ (see Figure S9 magnetoresistance transition at larger current windows). The origin of the current induced easy axis switching is Joule heating of the sample above $T_S$ (Figure S8b, Supporting Information). The main technological relevance for Joule heating induced switching would be in heat-assisted technologies such as Heat Assisted Magnetic Recording (HAMR), and comparison can be made with electrically controlled spin-oriented transitions demonstrated in magnetic semiconductors[8]. We note that the obtained switching current density is an order of magnitude lower for what has been calculated for achieving easy axis switching in proposed spin-reorientation assisted spin transfer torque devices[29], and for example, the current induced antiferromagnetic to ferromagnetic transition in FeRh devices[30].

## 2.3. X-ray spectroscopy probing of the spin-reorientation transition

In order to unravel the physics of the phase transition, X-ray Absorption Spectroscopy (XAS) and X-ray Magnetic Circular Dichroism (XMCD) measurements were carried out at the Co $L_{2,3}$ edges in Pt/Co(1.5 nm)/H$_2$Pc heterostructures. The XAS spectra indicate that some of Co atoms have a 2+ valence, as shown in Figure 3a, and this is the case for at all temperatures up to 300 K (see Figure S12, Supporting Information for room temperature XAS/XMCD spectra). The charged state was absent in reference films without molecules (Figure S13, Supporting Information). To rule out an effect due to oxidation, magnetic dichroism measurements were carried out on an oxidised sample (Figure S14). A comparison of dichroism signals of pristine and oxidised Co/H$_2$Pc interfaces at T = 2 K are shown in Figure S15. Our model estimates 15% Co$^{2+}$ at the surface of the molecule hybridised pristine sample (with 0 eV 10Dq value). We find that in the case of oxidised sample, both the XAS and the XMCD spectra are instead described by an entire Co$^{2+}$ layer, modelled with a 10Dq value of 0.7 eV (see Supporting Information Figure S16). The positively charged state and reduced moment are supported by DFT calculations (see Supporting Information, Table S1) with the electron depletion persisting several Co layers below the organic molecule. The strong π-d orbital hybridisation and

chemisorption[11] may be enabled by the ultra-high vacuum (10⁻¹⁰ mbar) during the molecule
sublimation on the metal surface.

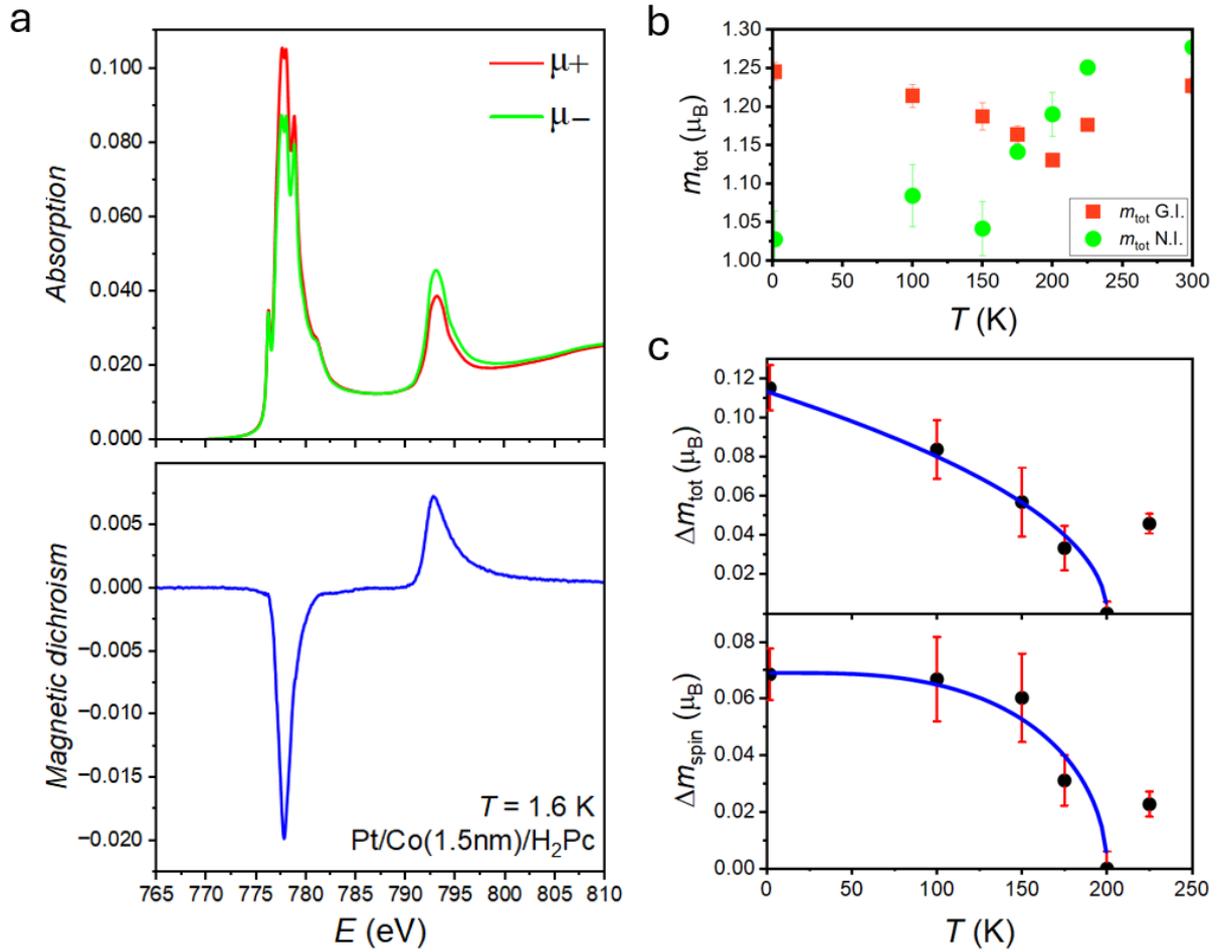

**Figure 3. XAS and XMCD measurements in a finite saturating field on a Pt/Co(1.5 nm)/H₂Pc structure. a**,
Circular polarisation dependent X-ray absorption spectrum and the corresponding magnetic dichroism
measured at Co edge at 2 T in grazing incidence, showing Co to be in 2+ state due to molecule chemisorption.
**b**, Extracted total magnetic moment in grazing (GI) and normal incidences (NI) where a planar magnetisation
onset below T=200 K is evidenced as GI total moment overtakes NI. **c**, Change in total moment and spin
moment in planar orientation fitted with scaling fits of $(T - T_C)^{1/2}$ and $tanh\left(1.74\sqrt{\left(T_C/T - 1\right)}\right)$,
respectively, with $T_C$ being 200 K.

The spectra were analysed using sum rules[31] as shown in Figure 3b and c. (see Figure S17 for sample
integrated XAS/XMCD spectra curves). Further to magnetometry, the easy axis switching can also be
observed in the total magnetic moment of Co ions, $m_{tot}$, derived from XMCD measurements in
normal (OOP) and grazing (planar) incidence (Figure 3b). The in-plane $m_{tot}$ overtakes the
perpendicular $m_{tot}$ below 200 K. Evidence of such magnetisation axis switching in extracted $m_{tot}$ is
absent on degraded films which is a further finding against oxidation. (see Figure S20, Supporting
Information). Figure 3c shows the change in planar $m_{tot}$ and spin moment, $m_{spin}$ from the
parameters transition onset point at 200K. The $(T - T_C)^{1/2}$ scaling relation predicted for frustrated
Kondo spin lattice systems[32,33] is found to describe $\Delta m_{tot}$ as a function of temperature. The change in $m_{spin}$ was found to be well described by a weak coupling many-body fit,

$$tanh\left(1.74\sqrt{\left(\frac{T_c}{T}-1\right)}\right).$$ The physical mechanism behind the phase transition is likely correlated with long-range spin interactions between the hexagon-pentagon hybridisation sites[12], as previously discussed in STM studies of metallo-molecular interfaces[17,19,34].

## 2.4. THz time-resolved magneto-optical Kerr effect (MOKE) spectroscopy of the spin-reorientation transition

The sharp nature of the spin-reorientation transition suggests possible applications in ultrafast devices, such as heat assisted memory systems, where switching times below 1 ns are needed[7,35]. We measured the timescales involved in the easy axis switching using time-resolved magnetooptical Kerr spectroscopy (TrMOKE). The magneto-optical Kerr effect (TR-MOKE) transients measured upon excitation with a 50 fs laser pulse (see Methods for more detail and Supporting Information, Figure S21 for the set up schematic) with the magnetic field and laser beam nearly perpendicular to the heterostructure plane are shown in Figure 4a. As depicted in Figure 4e, the initial demagnetisation of the sample is observed on a sub-picosecond timescale, corresponding to the negative part of the TR-MOKE transient. Near the spin reorientation transition, 230 to 290 K as shown in Figures 4b and c, the positive part of the TR-MOKE transient indicates an increase of the OOP magnetisation with a rise time of a few tens of picoseconds (≈40 ps at $T$ = 260 K with a weak excitation fluence of 0.12 mJ/cm²). The onset OOP magnetisation was found to decay exponentially with a sub-300 ps decay constant (see Figure S22, Supporting Information, for field dependence at a given temperature). Further data on the dependence of the light induced transient OOP magnetisation on magnetic field and excitation fluence is shown in Figure 4d. The data indicate that at a moderate excitation fluence of 0.5 mJ/cm² OOP magnetisation onset of ~40% can be achieved (with respect to the static saturated value), exceeding 50% at 1.5 mJ/cm². The rise-time of the OOP magnetisation however, which increases with increasing excitation fluence (see Figure S23, Supporting Information) is ~100 ps at 0.5 mJ/cm². Our data suggests a parameter of space for novel heat assisted technologies, where optimisation could be carried out to achieve optically induced spin-reorientation transition at remarkably low fluences ranging down to 0.12 mJ/cm². This is an order of magnitude smaller than those firstly used in HAMR systems for example, to write/erase data[36], secondly all optical helicity independent switching systems to switch magnetisation direction[37–41], and lastly magnetic tunnel junction structures demonstrating spin-reorientation transitions[42].

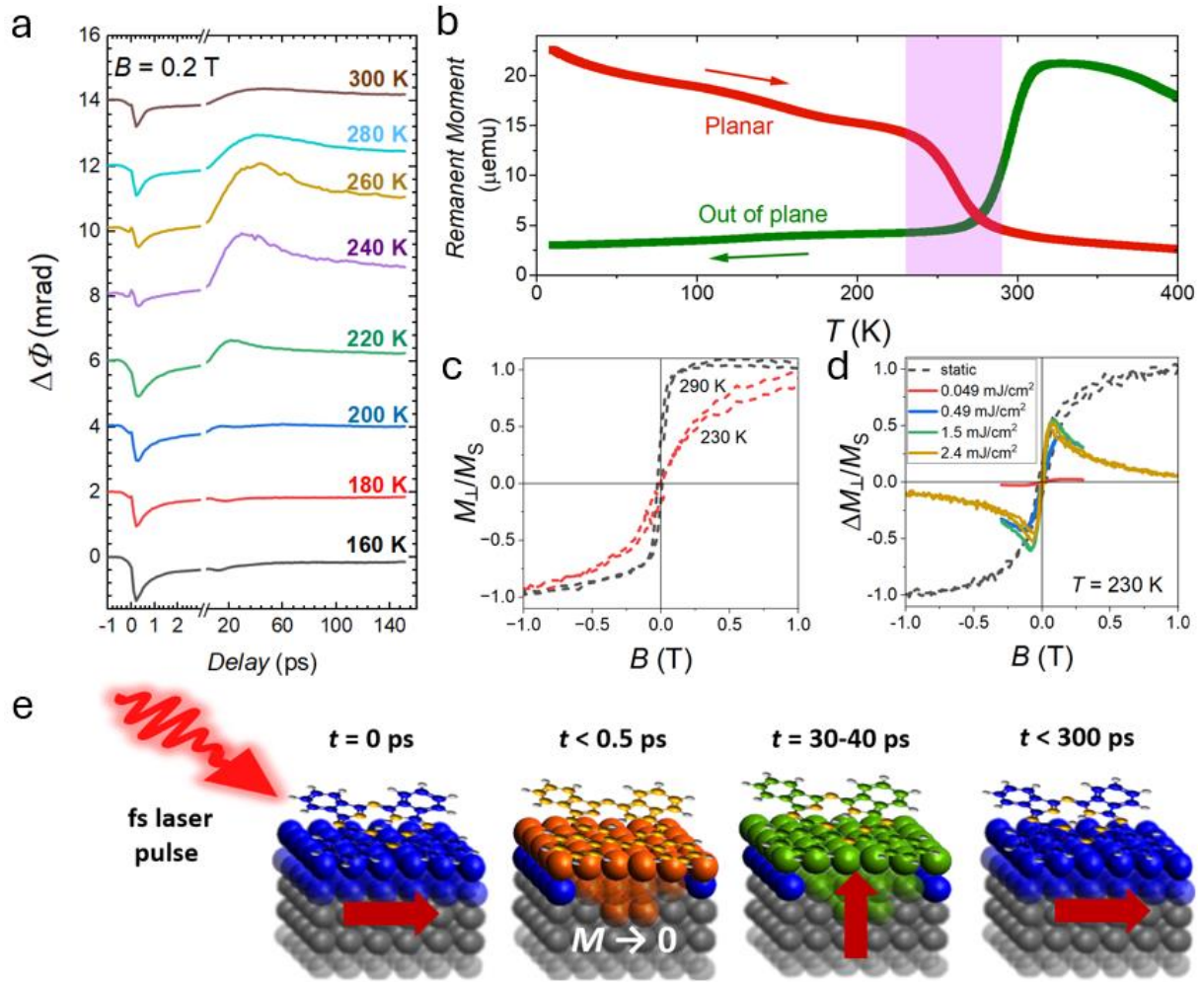

**Figure 4. THz magneto-optical Kerr effect (MOKE) spectroscopy of the spin-reorientation transition at Pt/Co(1.4 nm)/H₂Pc structure. a**, Temperature dependent polar time-resolved Kerr angle transients corresponding to the out-of-plane magnetisation measured at an excitation fluence of 0.12 mJ/cm² showing an onset of optical-pulse-induced out-of-plane spin-reorientation transition between 220 and 280 K. **b**, Planar and perpendicular magnetic remanence curves measured on the metallo-molecular structure with the spin-reorientation transition temperature window (from 230 K to 290 K) highlighted in pink. **c**, The out-of-plane static polar MOKE hysteresis around the transition window at 230 K and 290 K, showing the easy axis emergence on the latter. **d**, The normalised field-dependent magnitude of the transient out-of-plane magnetisation, $\Delta M_\perp$, in comparison to the static out-of-plane magnetisation, $M_\perp$, at the optimal $T$ = 230 K as a function of the excitation fluence specified in the legend. **e**, Schematic illustrating the picosecond spin-reorientation transition dynamics (suggested in **a**) with the planar magnet demagnetisation below 0.5ps followed by the spin-reorientation transition after the laser-induced heating on 30-40 ps timescale, and planar magnetisation re-emergence below 300 ps.

## 3. Conclusion

In conclusion, a low-energy, ultrafast unconventional spin-reorientation transition has been observed near room temperature in Pt/Co/molecular heterostructures with competing magnetic interactions. The transition onset and window are controlled by the ferromagnet thickness and molecule used. For appropriate ferromagnetic film thickness range, the molecular interface coupling induced in-plane magnetisation dominates at low temperatures, whereas the perpendicular magnetisation arising from spin orbit coupling at the Pt interface is dominant at higher temperatures. The magnetisation easy axis direction is therefore not fixed by the structure composition but can be rotated from in-plane to out-of-plane with ultra-low power, paving the way to low-power information storage and memory applications. The electric switching of the magnetisation easy axis is demonstrated to be possible via local heating with a current density that is an order of magnitude lower ($\approx$ $1 \times 10^5$ A/cm$^2$) than those used to e.g. introduce antiferromagnetic – to ferromagnetic transition in FeRh[30], or the densities calculated for spin-reorientation assisted spin-transfer torque system proposals[29]. In heat assisted technologies, optical or other local heating mechanisms can be used to switch such structures with a local temperature change of just 50 K. Compared to the existing HAMR systems, where approaching towards Curie temperatures is necessary, order of magnitude lower laser fluences of around 0.12 -0.5 mJ/cm$^2$ are shown to be enough for optically induced spin-reorientation transition as well as an operation timescale of sub 300 ps. Demonstrated lower laser fluence is also found to be order of magnitude lower than systems showing all optical helicity independent switching. In future work, further optimisation of energy and timescales should be possible by tuning substrate and film thicknesses, as well as studying other magnetic and molecular materials. By tuning Co thickness to higher values, spin-reorientation transition temperature can be tuned above room temperature creating a platform for practical memory devices. XAS/XMCD data measured at Co edge evidence the onset of anisotropy switching at the atomic level, where the in-plane magnetisation can be attributed to correlation of spin moments at the molecular interface and be described by a scaling fit. Beyond single ion Kondo effect measured in polycrystalline gold molecule interface films[43], the candidate physical mechanism behind the observed planar magnetic phase transition is a temperature dependent onset of a correlated spin lattice effect, arising due to a hybridised carbon supramolecular lattice of spins formed at the metallo-molecular interface in ultra-high vacuum[17,19,34,44].

## 4. Experimental Section

**Thin film growth and characterisation:** Thin-film structures were grown on 0.65 mm thick c-plane sapphire films. The (111) textured Pt layers of $\approx$4 nm thickness (Figure S24, Supporting Information, shows the XRD peak) were grown at 500 °C with e-beam evaporation at a growth rate of $\approx$0.1Å/s. The substrate was than cooled down to room temperature and the Co-layer was grown also with e-beam evaporation technique at a rate of $\approx$0.1 Å/s. Within the same chamber, organic molecule layers were sublimed onto the Co surface at a pressure of $\approx 5 \times 10^{-10}$ mbar with a rate of $\approx$0.3 Å/s and until a thickness of 20 nm was measured through a quartz monitor. The cap layer was magnetron sputtered on top of the thin film structure, with the material used for capping being Cu with a thickness of $\approx$ 15 nm. Films were then structurally characterised using x-ray reflectivity and x-ray diffraction (Figure S25, Supporting Information). Transmission electron microscopy characterisation of similar structures can be found in elsewhere[45,46]. Raman characterisation of a phthalocyanine metallo-molecular structure is shown in Supporting Information, Figure S26.

**Magnetometry:** Magnetisation measurements were carried out using SQUID-magnetometer (MPMS Q.D.) which offers a resolution over $10^{-8}$ emu accompanied by temperature control.

**XAS/XMCD:** X-ray absorption and magnetic circular dichroism measurements were carried out at BL-29 BOREAS of ALBA synchrotron light source[47]. For XAS studies, a thinner Cu 3nm was deposited on the metallo-molecular structures to prevent attenuation by the Cu cap. The error bars were obtained on magnetisation values determined post sum rule analysis by taking the range of moment values obtained post subtraction of varying values of background signal. The $Co^{2+}$ XAS and XMCD calculations were carried out by crystal field multiplet (CFM) simulations using CTM4XAS[48]. We have chosen 10Dq=0 eV assuming the Cobalt is in spherical symmetry. For calculations, the final Slater integrals, F($pd$), G($pd$), and F($dd$) were considered a standard reduction of 80% to their Hartree-Fock values. The CFM simulations were done at $Co^{2+}$ electronic configuration and simulated for the $C_4$ crystal symmetry. The $M$ value was chosen as 6 meV while the $D_t$ and $D_s$ remain to zero. XMCD simulations for Co/H2Pc interface were carried out using an 85% composition of Co, and 15% composition of $Co^{2+}$ with a 10Dq value of 0. For modelling oxidised sample, CoO spectrum with the crystal field of 10Dq $\sim$ 0.7 eV was found to match to the experimental spectra.

**Electronic transport:** 4-probe resistance measurements were carried out in continuous flow cryostats on devices of narrow cleaved strips of films, with a Keithley 6621 DC and AC current source with a Keithley nanovoltmeter being used for current induced switching measurements.

**Time resolve magneto-optical Kerr effect (TrMOKE):** Measurements were conducted by means of a two-colour pump-probe setup based on a high repetition-rate (250 kHz) 50-femtosecond Ti:sapphire laser amplifier and a split-coil superconducting 7 T optical magnet with a variable temperature He exchange gas sample insert (see Supporting Information, Figure S21). A part of the output pulse train was used to derive the pump pulses that were either at the laser fundamental (λ=800 nm, 1.55 eV) or frequency doubled (λ=400 nm, 3.1 eV photon energy). The probe pulses with a variable time delay were derived from the remaining fundamental pulse train. The reflected-probe-beam transient (and static) polarisation rotation was detected by means of a balanced detection using a Wollaston prism and a pair of silicon PIN photo diodes. The pump and probe beams were modulated at two different frequencies in a couple of kHz range with an optical chopper and a standard lock-in detection scheme was used to acquire the photodiodes sum and/or differential signal. The magnetic field and the laser beams were nearly perpendicular to the heterostructure plane in a polar TrMOKE configuration.

**Density functional theory:** All first-principles calculations were performed using the Vienna Ab Initio Simulation Package (VASP) within the projector-augmented wave (PAW) formalism. The exchange-correlation functional was described using the generalized gradient approximation (GGA) as formulated by Perdew, Burke, and Ernzerhof (PBE). To accurately capture on-site Coulomb interactions, we enabled the LDA+U (U=0, J=0) correction. A spin-polarized DFT calculation was performed using the PAW method in VASP. The Co substrate was modelled as a 4x6 (17.96 x 15.06 Å) 4-layer slab of the (111) surface of FCC Co with 15 Å vacuum in the c direction. A plane-wave kinetic energy cutoff of 400 eV was employed in a ferromagnetic spin configuration. The 2D Brillouin zone was sampled with a 2x2 k-point grid centered at the Gamma point. Structural relaxation was carried out with a force tolerance of 0.05 eV Å⁻¹. The RMM-DIIS algorithm was used for electronic minimization to accelerate convergence with total energy and force convergence criteria set to $10^{-4}$ eV and -0.05 eV/Å, respectively, ensuring an accurate determination of the electronic structure. To capture the magnetic properties of the system, spin polarization was explicitly included, and the initial magnetic moments were defined for Co atoms. To account for dispersion interactions, the DFT-D3 van der Waals correction was enabled. Bader analysis was performed on the all-electron charge density by partitioning it into atomic regions using the grid-based zero-flux method. The charge integration was performed using the Henkelman algorithm, ensuring an accurate evaluation of atomic charges.

## Supporting Information

Supporting information is available online.

## Acknowledgements

We thank the Engineering and Physical Sciences Research Council in the UK for financial support via the grants EP/S030263/1 and EP/X027074/1. We also acknowledge the support of the EC project INTERFAST (H2020-FET-OPEN-965046).

## Conflict of Interest

The authors declare no competing interests.

## Data Availability Statement

Data is available from corresponding authors upon reasonable request.

## Keywords

Magnetism, spin-reorientation transition, ultrafast switching, current induced switching, phase transition

# Supporting Information

## Low-energy, ultrafast spin reorientation at competing hybrid interfaces with tunable operating temperature

*Servet Ozdemir*[*], *Matthew Rogers, Jaka Strohsack, Hari Babu Vasili, Manuel Valvidares, Thabahb Haddadi, Parvathy Harikumar, David O'Regan, Gilberto Teobaldi, Timothy Moorsom, Mannan Ali, Gavin Burnell, Bryan J Hickey, Tomaz Mertelj, Oscar Cespedes*[*]

**Supporting Note 1**

In order to model the spin-reorientation transition, the usual thermal activated magnetic remanence model was combined with a logistic-function

$$\left(1 - exp\left(\frac{-E_a}{k_B T}\right)\right) \cdot \left(\frac{1}{1 + \exp(-k(T - T_S))}\right)$$

where one can immediately see that the steepness function $k$ defines the transition window, that we can re-write as a spin canting temperature window as

$$\left(1 - exp\left(\frac{-E_a}{k_B T}\right)\right) \cdot \left(\frac{1}{1 + \exp\left(-\frac{(T - T_S)}{T_w}\right)}\right)$$

with both quantities $T_S$ and $T_w$ obtainable through a fit of the model to the experimentally observed remanence curves.

**Supporting Information Tables**

| Co -layer | Total $m$ ($\mu_B$) | $m$ ($\mu_B$) Per Co atom | Total $\Delta Q$ (e) | $\Delta Q$ (e) per Co atom | Change in PAW integrated atomic charges (e) |
|---|---|---|---|---|---|
| 1 | 80.4085 | 1.6750 | -0.6051 | -0.013 | +0.725 |
| 2 | 76.3707 | 1.5910 | +0.6109 | +0.013 | +0.523 |
| 3 | 75.7295 | 1.5776 | +0.1409 | +0.003 | +0.531 |
| 4 | 73.0390 | 1.5216 | +3.8517 | +0.080 | +0.624 |

**Table S1**: VASP derived, layer resolved analysis of (PAW-core integrated) magnetic moments (column 2 and 3), change in atomic charges (column 4 and 5) and change in total Bader derived (or PAW integrated) atomic charges (column 6). The sign convention used means a positive value of change in net charge in columns 4-6 means charge depletion. Layer 4 is the topmost $H_2Pc$ molecule hybridised layer.

| Interface molecule | Switching Temperature |
|:---:|:---|
| CoPc | 245 ± 5 K |
| $C_{60}$ | 287 ± 5 K |
| CuPc | 305 ± 5 K |
| $H_2Pc$ | 315 ± 5 K |

**Table S2**: Extracted switching temperatures on Pt/Co(1.7nm) interfaces using the phenomenological model described above.

**Supporting Figures**

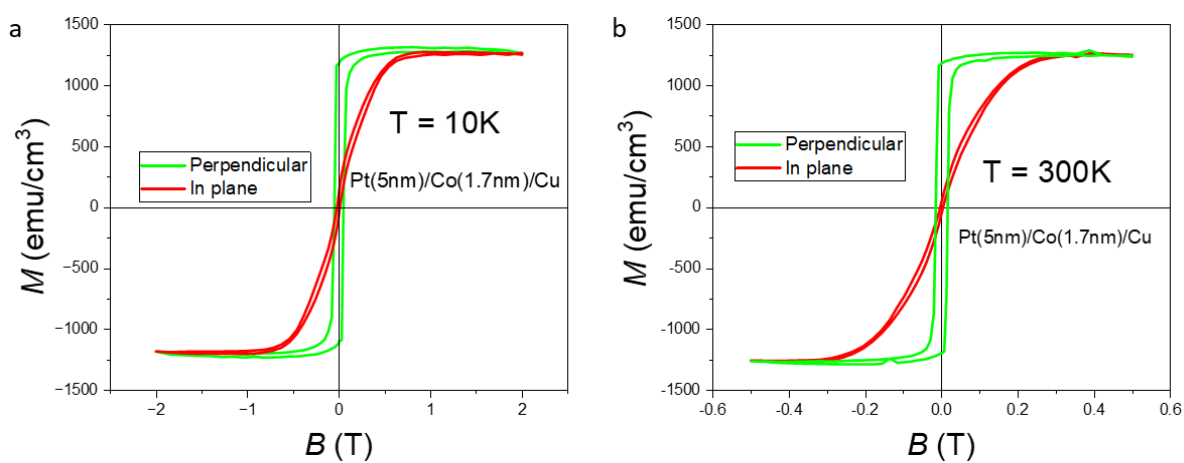

**Figure S1.** Perpendicular magnetic anisotropy reference samples without molecules. Pt/Co(1.7nm)/Cap film measured at perpendicular and planar orientations at **a,** T=10K, and **b,** T= 300K, **b**.

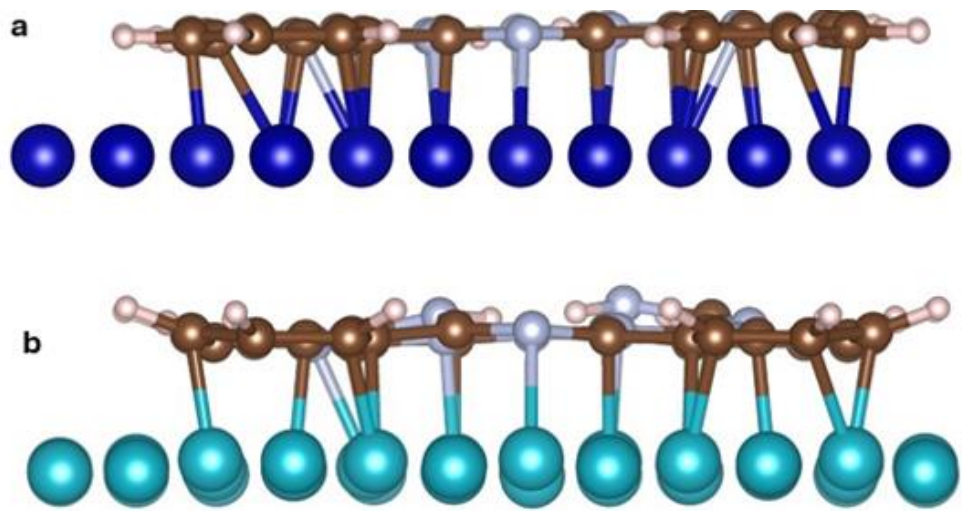

**Figure S2.** Flat lying Pc molecule depicted on Co(111) surface. **a**, initial metal-molecule contact leading to hybridisation **b**, post hybridisation relaxation showing the Co atom distortions.

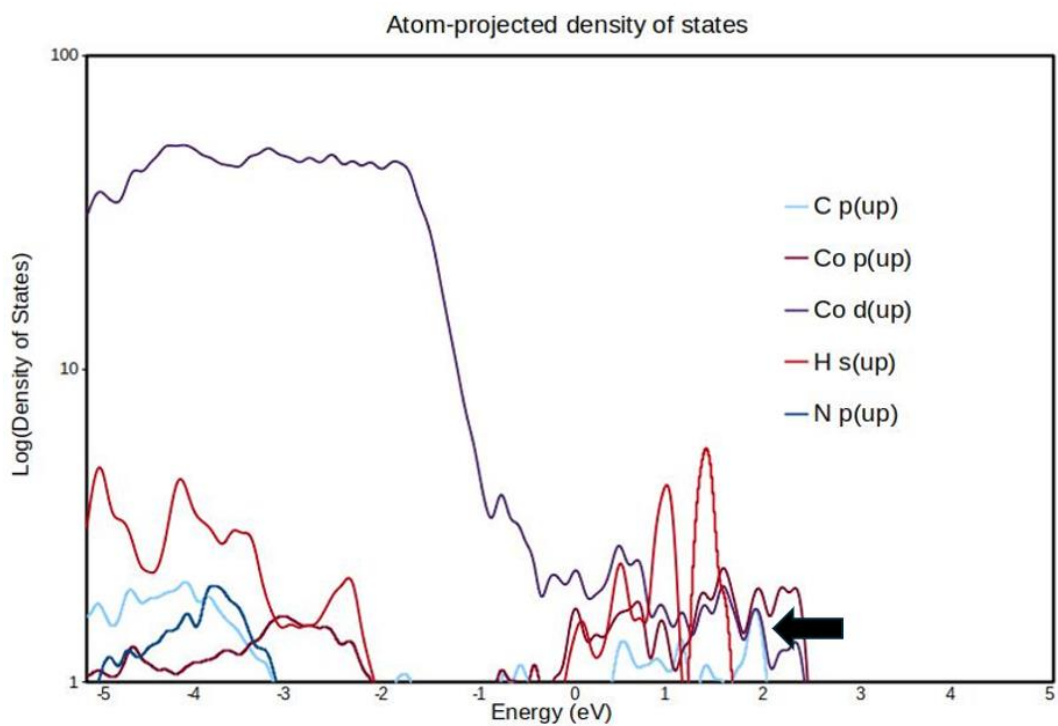

**Figure S3.** Atom projected density of states at the $H_2Pc$ and Co (111) interface. The region of overlap between Co d and C p orbitals can be seen pointed by the arrow suggesting hybridisation between the respective atoms at the interface.

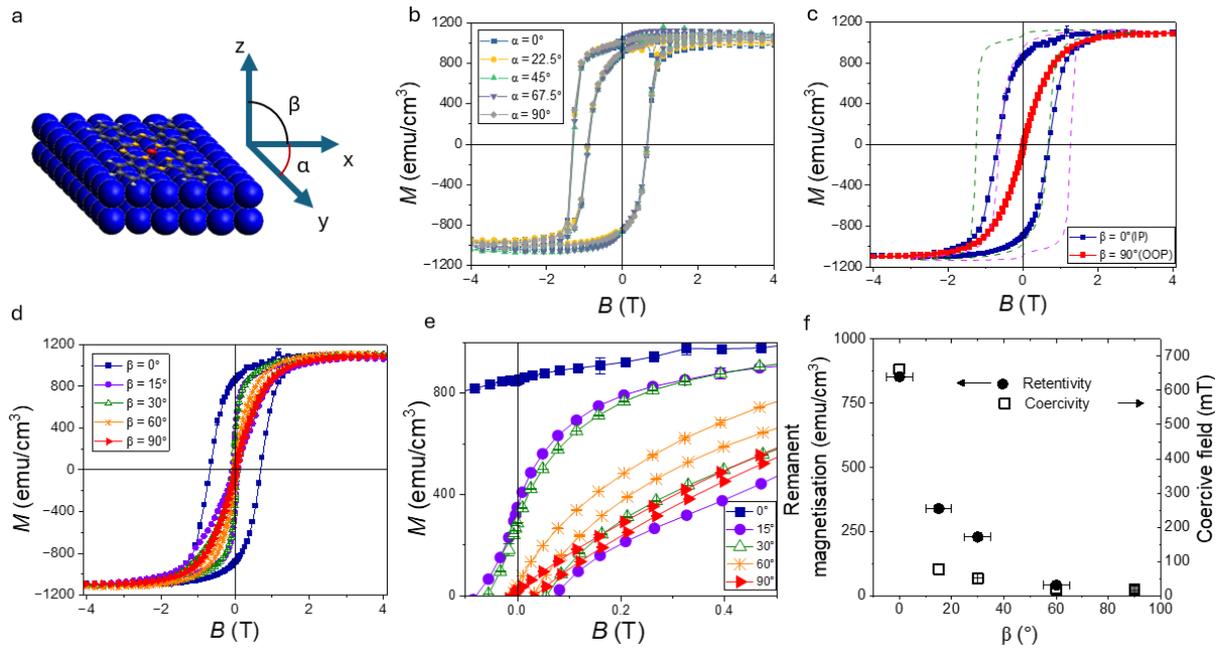

**Figure S4.** Low temperature (*T*= 10K) magnetic field rotation experiments for magnetic hysteresis on a Pt/Co(1.7nm)/C60 interface. **a,** Thin film schematic and the respective 3 axis labelled x, y and z, with in plane magnetic field rotation angle labelled α, and out of plane magnetic field rotation angle labelled β. **b,** Magnetic hysteresis curves measured at varying in plane rotation angles, α, showing no effect of in-plane rotation on remanent magnetisation and coercive field. **c,** Hysteresis loops measured at different two out of plane rotation angles β = 0° and β=90° suggesting in-plane magnetic anisotropy suggested by higher coercivity and larger remanent magnetisation at zero field for β = 0°(dashed lines correspond to magnetic pinning induced on the sample by field cooling at opposite polarities). **d,** Out of plane magnetic field rotation angle, β, dependence of magnetic hysteresis curves, showing an easy to hard axis switching as the field is rotated from β = 0° to β = 90°. **e,** Magnified version of graph presented on panel d, where smaller remanent magnetisation at zero field is evident at increasing angles β, as field is switched from planar (0°) to perpendicular (90°) orientation. **f,** Remanent magnetisation (retentivity -left y axis) and coercivity (right y axis) as a function of angle β, confirming magnetic anisotropy of the sample to be in-plane when β = 0°.

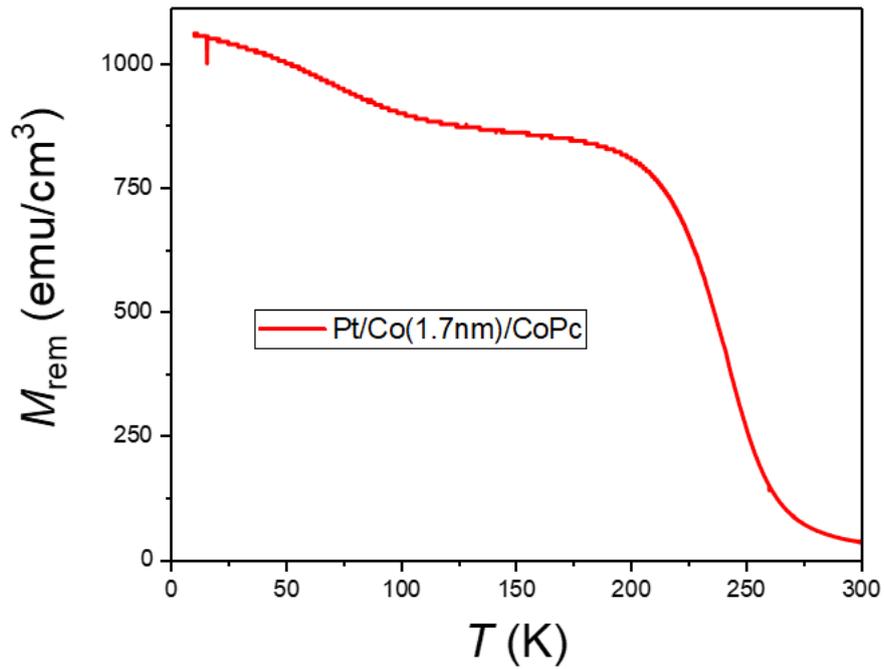

**Figure S5.** Remanent magnetisation measured post field cooling at 2 T, as a function of temperature during warming from 10K to 300K on Pt/Co(1.7nm)/CuPc heterostructure.

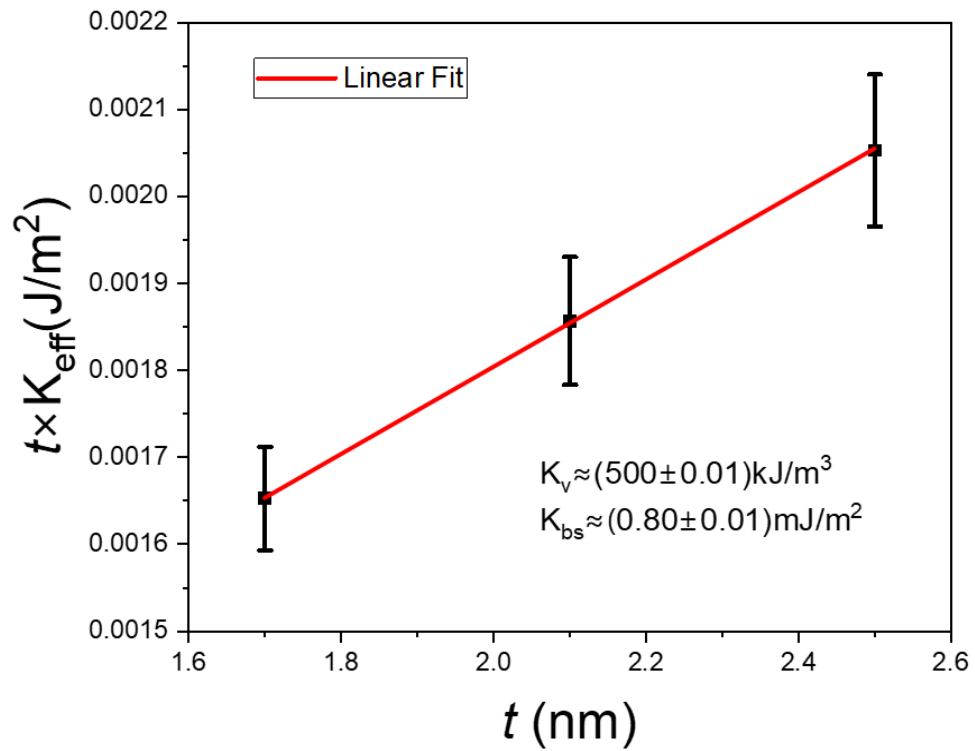

**Figure S6.** Pt/Co interface anisotropy constant estimation. Varying Co thickness hard axis extracted $K_{eff}$, multiplied by $t_{Co}$ plotted as a function of $t_{Co}$ yielding the volume and bottom interface anisotropy energies with slope and intercept respectively.

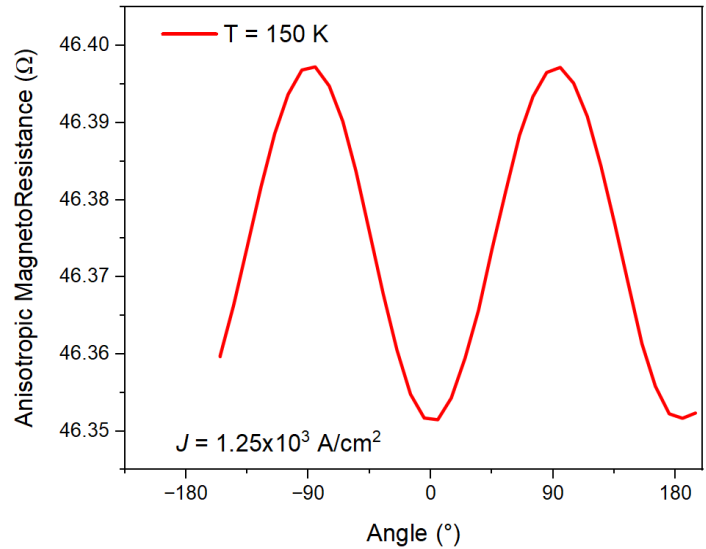

**Figure S7.** Anisotropic magnetoresistance curve measured at 150 K at a 3T field as magnetic field is rotated with respect to current (0 to 180 degrees), with the resistance minima corresponding to angles at which current is aligned to magnetic field.

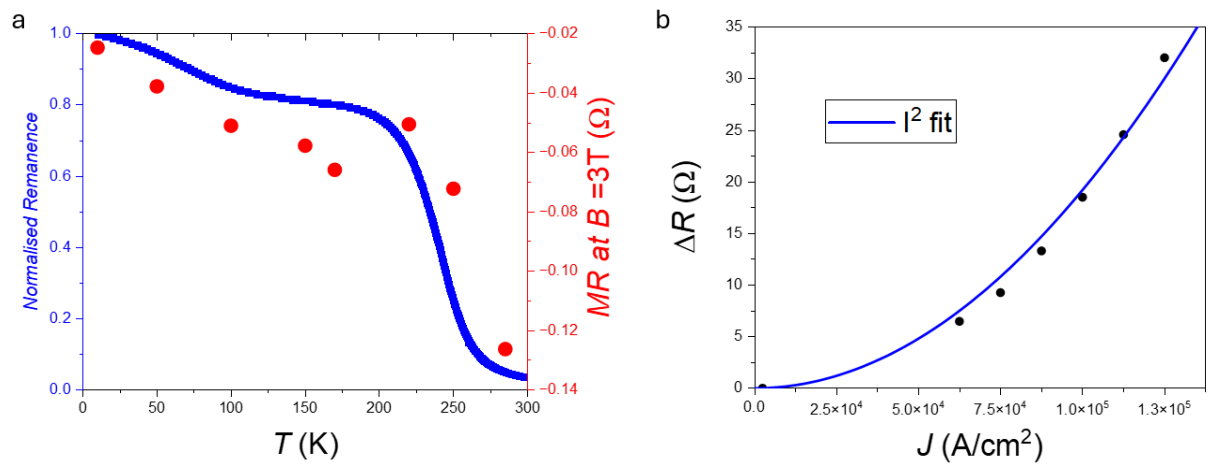

**Figure S8.** Enhanced negative magnetoresistance post spin-reorientation transition and Joule heating of the sample on Pt/Co(1.7nm)/CoPc interface. **a**, Planar normalised remanence (blue) and magnetoresistance measured at 3T (red, MR at $B$ = 3T) planar magnetic field aligned to current (0°) as function temperature, showing an enhancement in negative magnetoresistance as easy axis is switched from in plane to out of plane. **b**, Joule heating fit (blue) to the variation in ΔR with ΔR being the change in zero magnetic field resistance due to heating at high current densities.

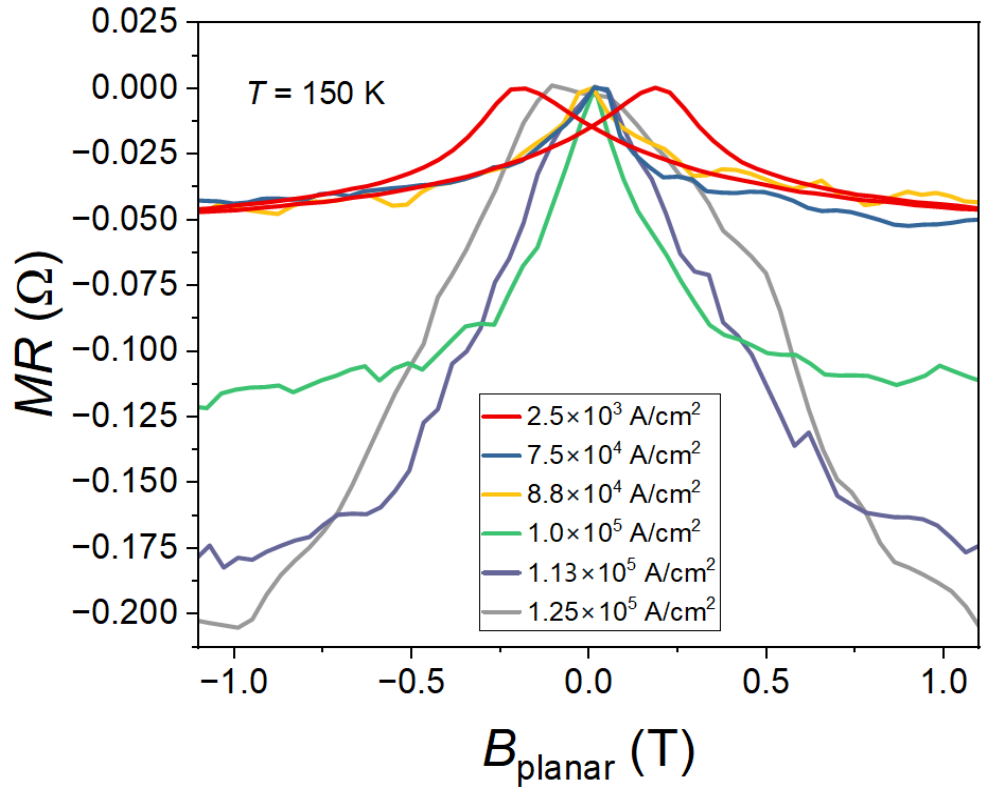

**Figure S9.** AMR magnetic field sweeps carried out at different current densities below switching temperature, showing magnetoresistance corresponding to current densities below a critical switching current ($2.5 \times 10^3 \ A/cm^2$ -red, $7.5 \times 10^4 \ A/cm^2$ -blue, $8.8 \times 10^4 \ A/cm^2$- yellow), and above a critical switching current ($1 \times 10^5 \ A/cm^2$-green, $1.13 \times 10^5 \ A/cm^2$-purple, $1.25 \times 10^5 \ A/cm^2$-grey). The jump in magnetoresistance at 1T is 0.068 Ω, 0.135 Ω, 0.159 Ω at current windows of $\pm 1.2 \times 10^4 \ A/cm^2$, $\pm 2.5 \times 10^4 \ A/cm^2$, $\pm 3.7 \times 10^4 \ A/cm^2$ respectively.

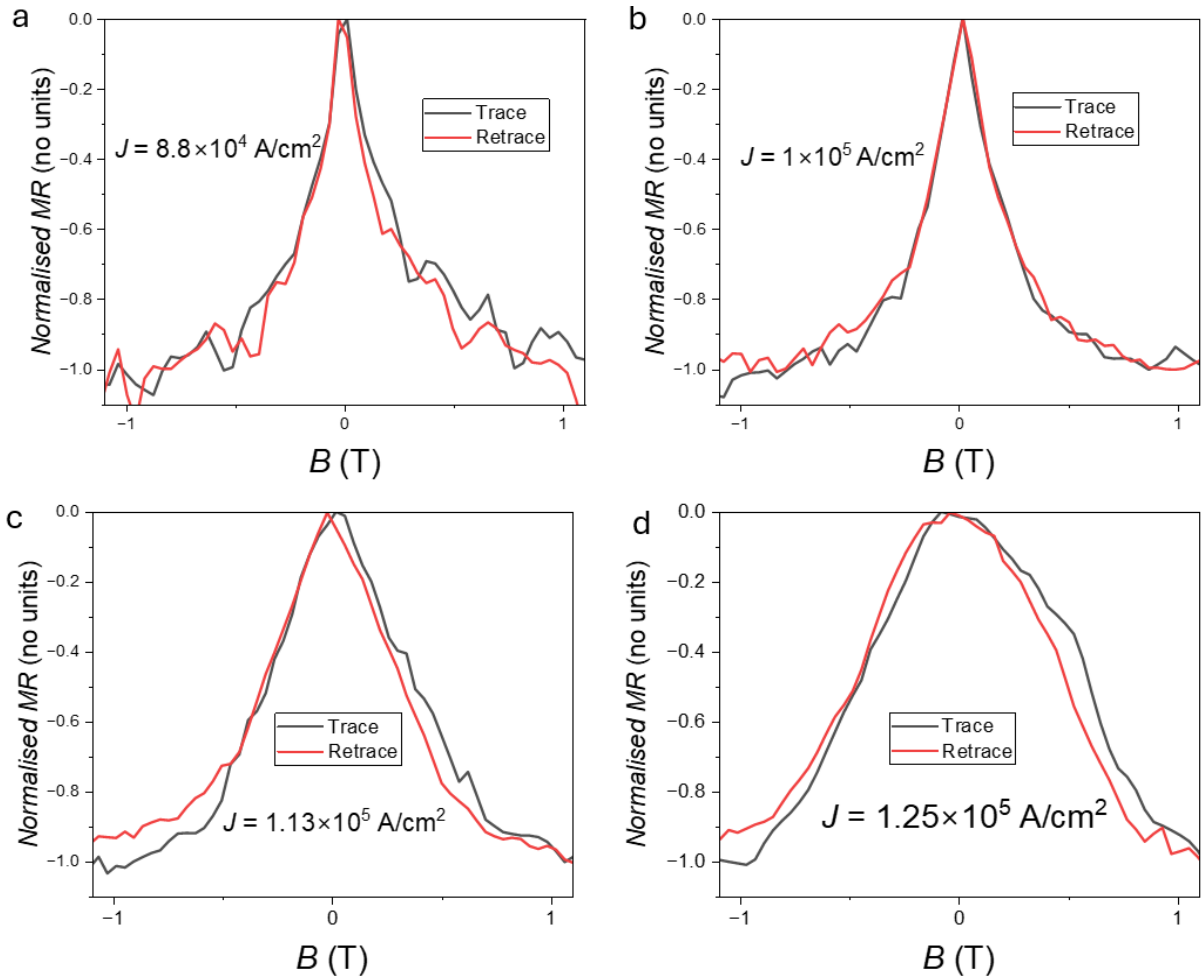

**Figure S10.** Normalised magnetoresistance Trace -Retrace curves measured at T=150 K with various high currents **a**, $8.8 \times 10^4$ A/cm² , **b,** $1 \times 10^5$ A/cm² **c**, $1.13 \times 10^5$ A/cm² and **d,** $1.25 \times 10^5$ A/cm² showing absence of butterfly-like magnetoresistance, when a Joule heating induced spin-reorientation transition takes place.

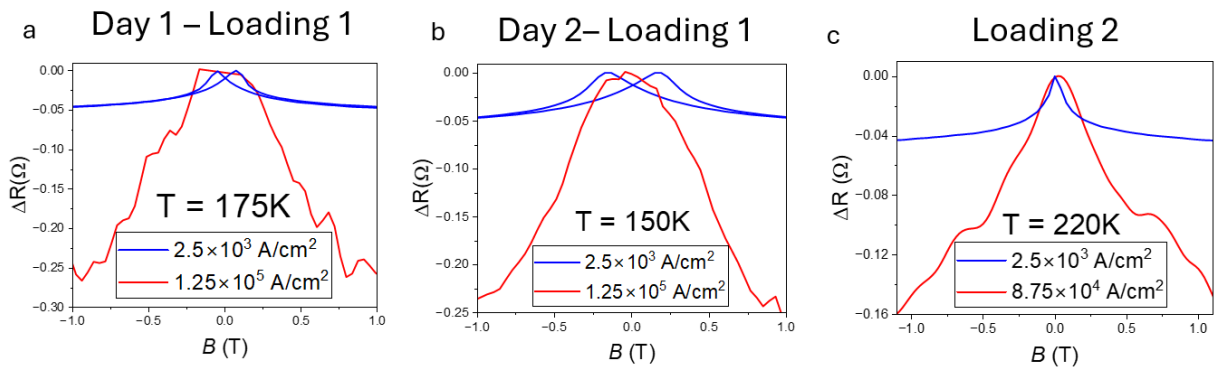

**Figure S11.** AMR magnetic field sweeps carried out at different current densities below switching temperature, showing magnetoresistance corresponding to planar (blue), perpendicular (red) magnetised states, measured at different days in a given cryostat loading **a** and **b**, and measured at a different cryostat loading **c**, suggesting stability of the devices post Joule heating cycles.

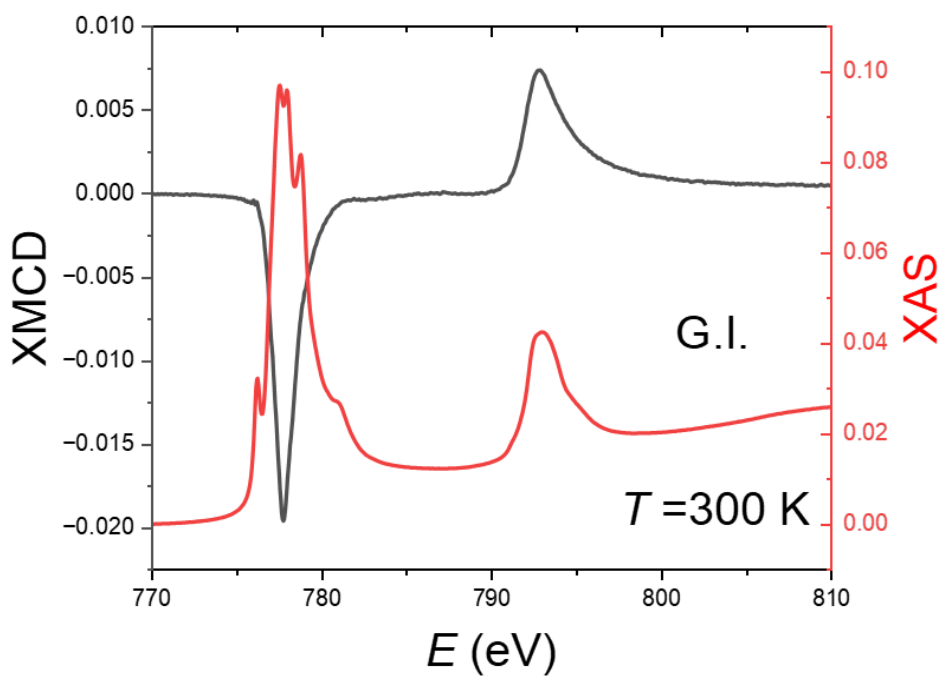

**Figure S12.** XMCD (black) and XAS (red) spectrum of Pt/Co(1.5nm)/H$_2$Pc sample at room temperature with the XAS spectrum suggesting Co to be in 2+ state.

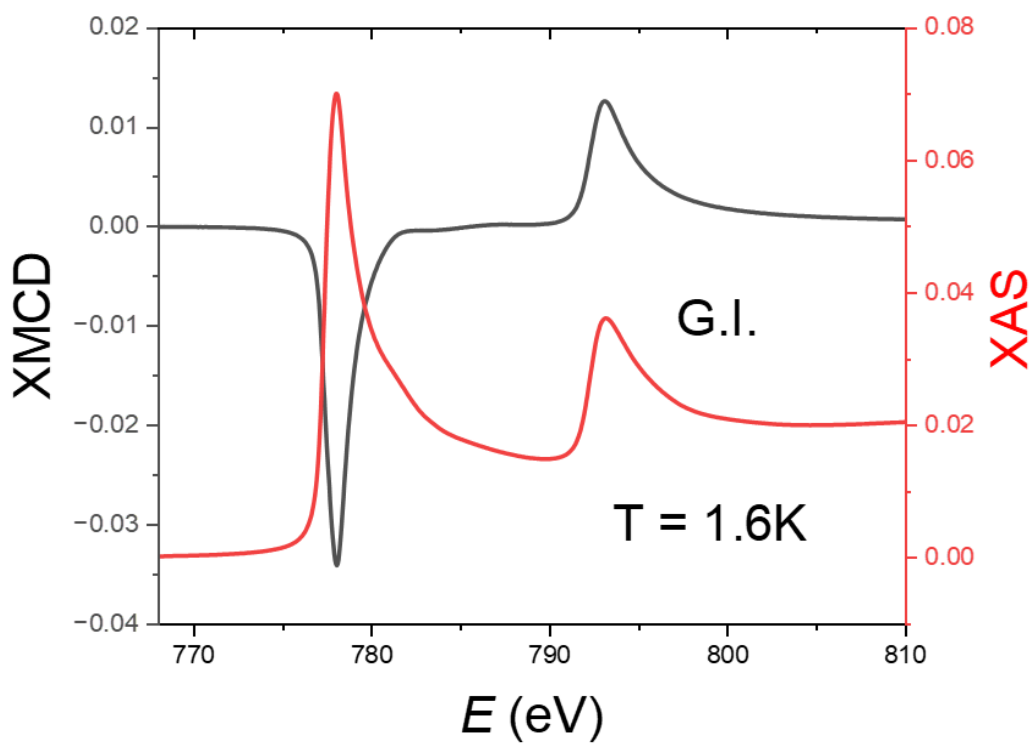

**Figure S13.** XMCD (black) and XAS (red) spectrum at grazing incidence measured on the capped Pt(5nm)/Co(1.5nm) reference sample, with the ionised Co 2+ state shown to be absent.

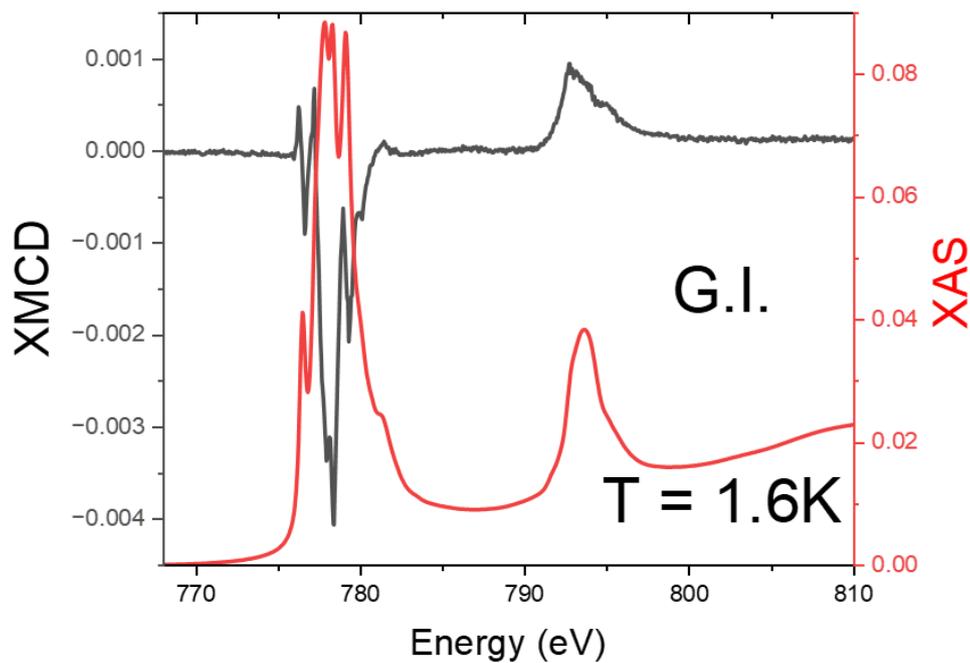

**Figure S14.** XAS and XMCD spectrum of improperly capped Pt/Co(1.5nm)/CuPc interface. Disrupted magnetism and an altered Co2+ state due to oxidation is evident.

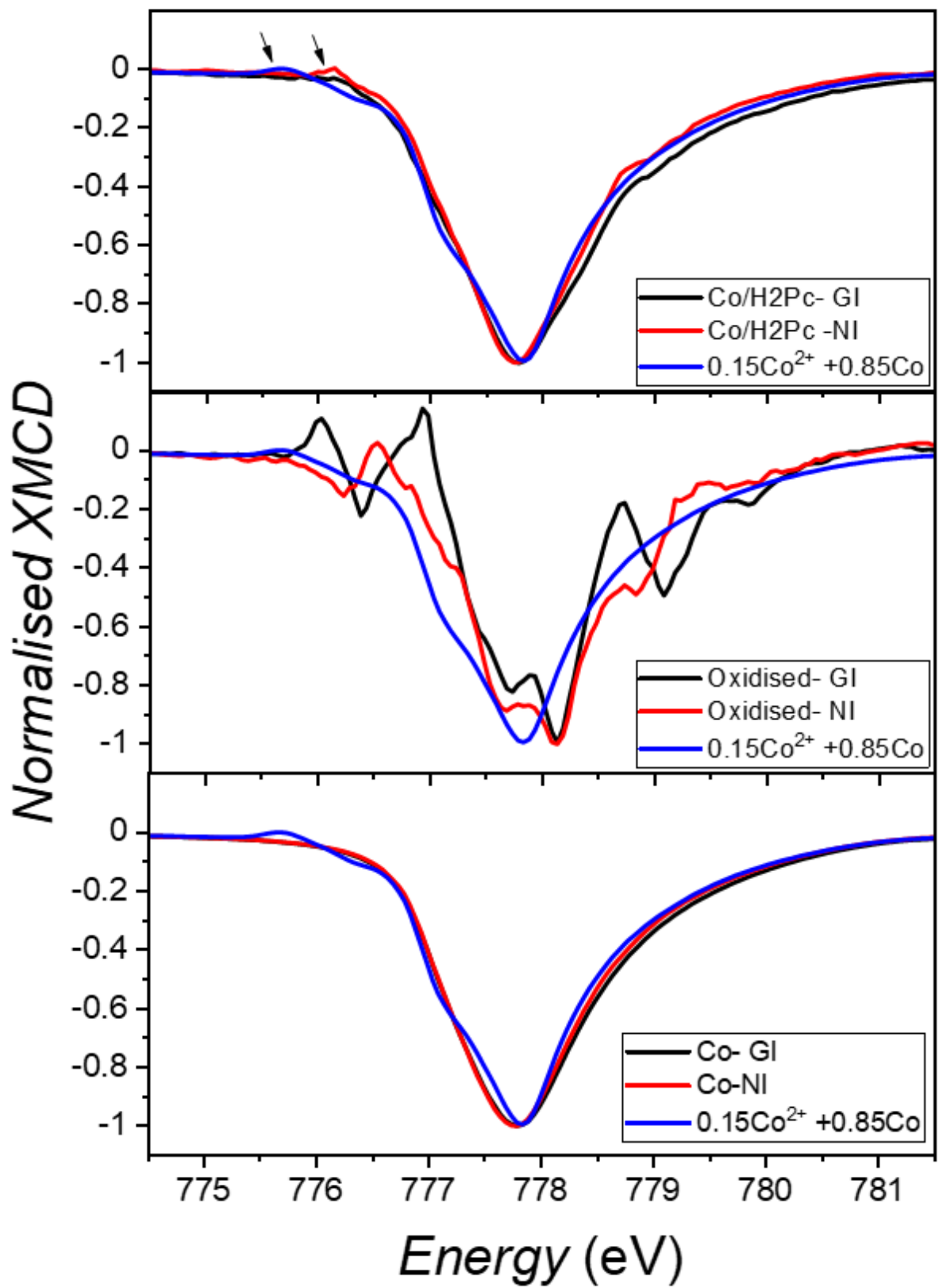

**Figure S15.** Normalised XMCD data (black-G.I., red-N.I.) on Co/H2Pc, oxidised Co, and reference Co on top, middle and bottom graphs respectively plotted with a calculation (blue) that models XMCD spectrum of a 15% $Co^{2+}$ and 85% Co sample, with the 10Dq value used for $Co^{2+}$ being 0 eV.

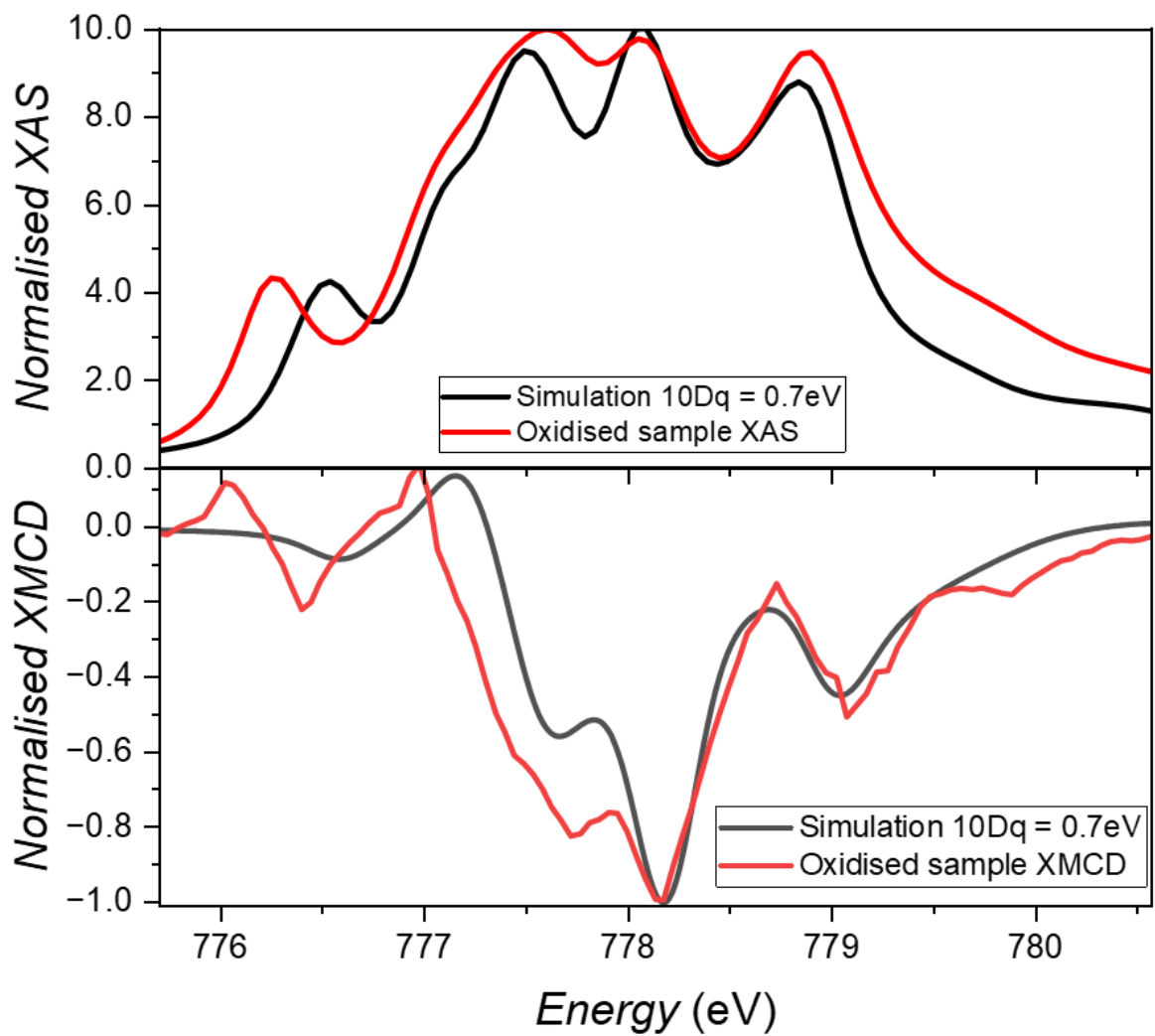

**Figure S16.** Normalised experimentally measured XAS and XMCD (red curves, top and bottom graphs respectively) on an oxidised sample, plotted with XAS and XMCD CoO simulations (black curves, top and bottom graphs respectively) calculated using a 10Dq value of 0.7eV on an entirely $Co^{2+}$ sample.

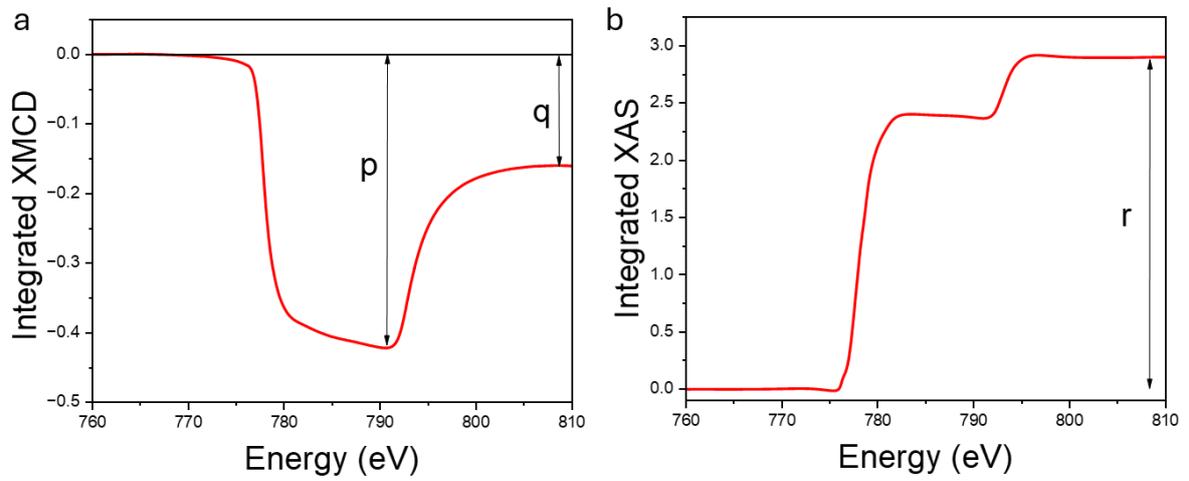

**Figure S17.** Integration of **a**, XMCD and **b**, XAS curves measured at grazing incidence at T=1.6K at a magnetic field of 2T on a Pt/Co(1.5nm)/H₂Pc structure, showing determination of sum rule quantities p, q and r.

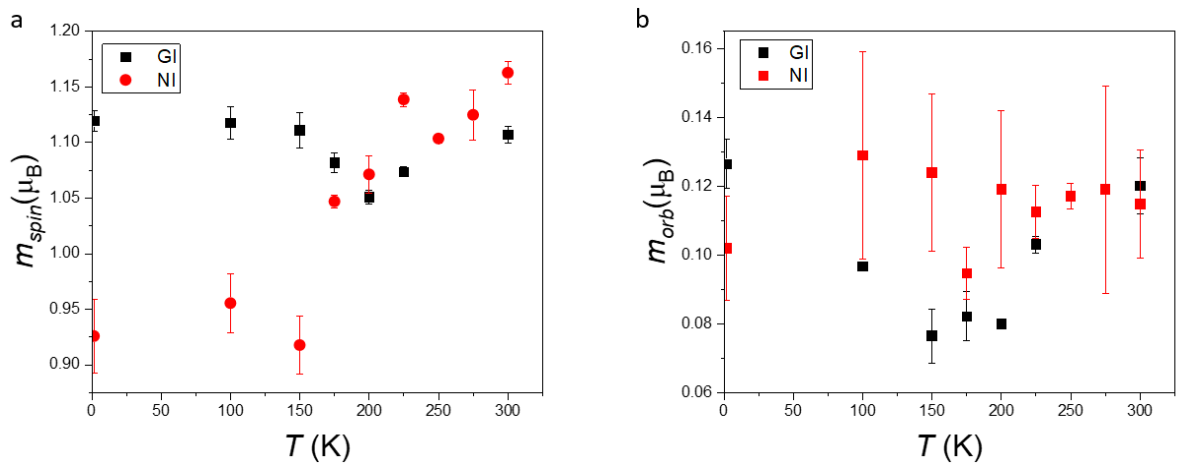

**Figure S18.** Further data on Pt/Co(1.5nm)/H2Pc interface reported in the main text. Extracted **a**, spin and **b**, orbital moment components for the Pt/Co(1.5nm)/H2Pc interface.

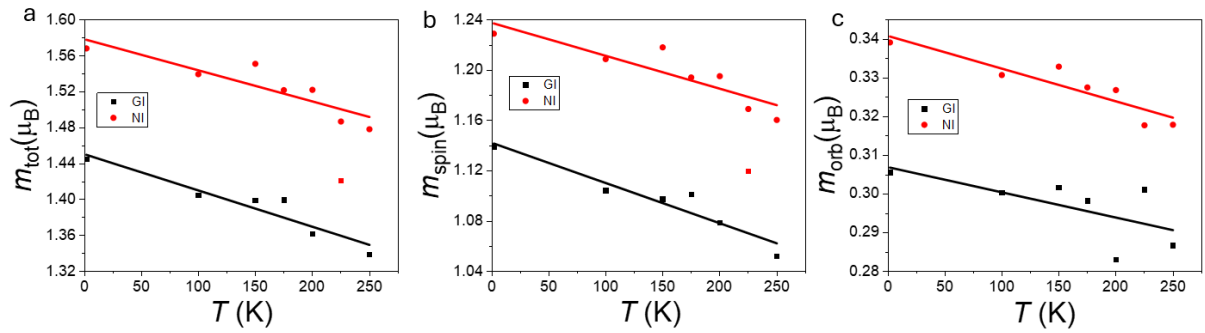

**Figure S19. a, b, c,** extracted total moment, spin moment and orbital moment on the reference Pt/Co(1.5nm)/Cap sample showing perpendicular magnetic anisotropy at all temperatures.

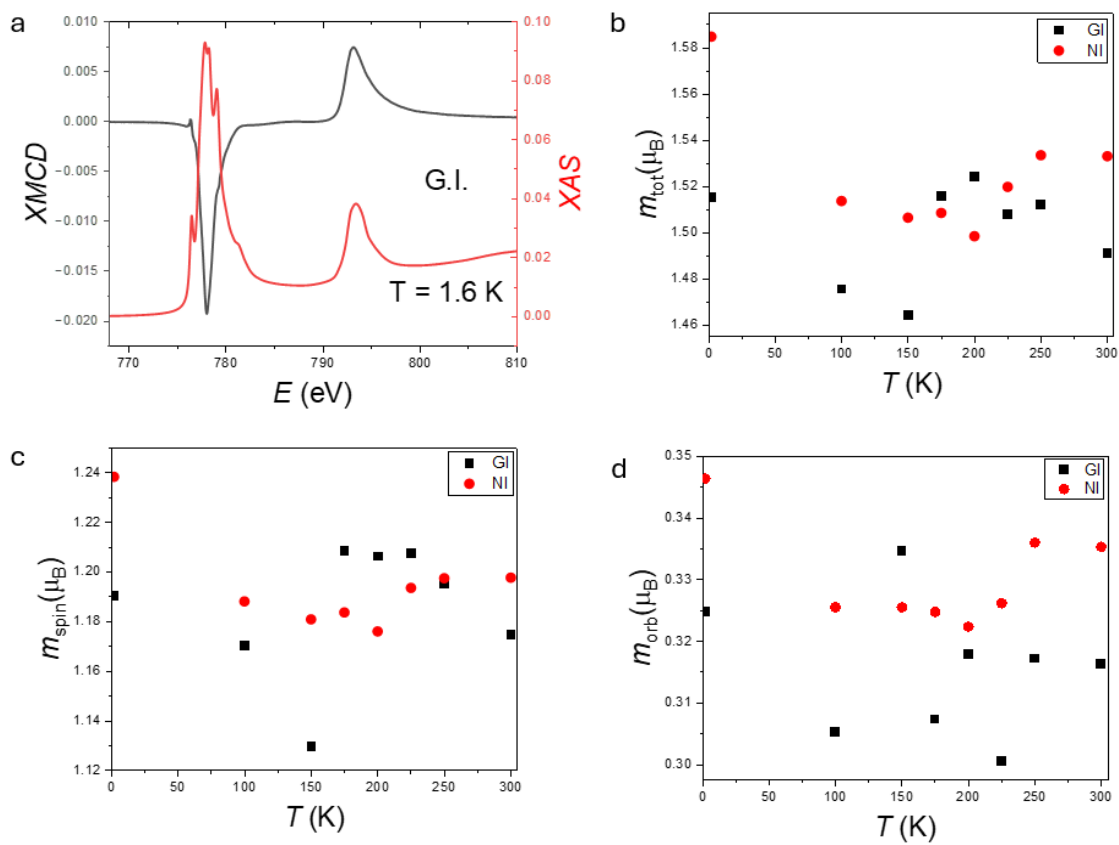

**Figure S20. a,** XAS and XMCD data on a Pt/Co(1.5nm)/CuPc/Cu (3nm) structure measured a year after growth. **b, c, d,** Total magnetic moment, spin moment and orbital moments respectively, with temperature dependence of total magnetic moment showing lack of grazing incidence magnetisation overtaking normal incidence, spin reorientation transition therefore being absent.

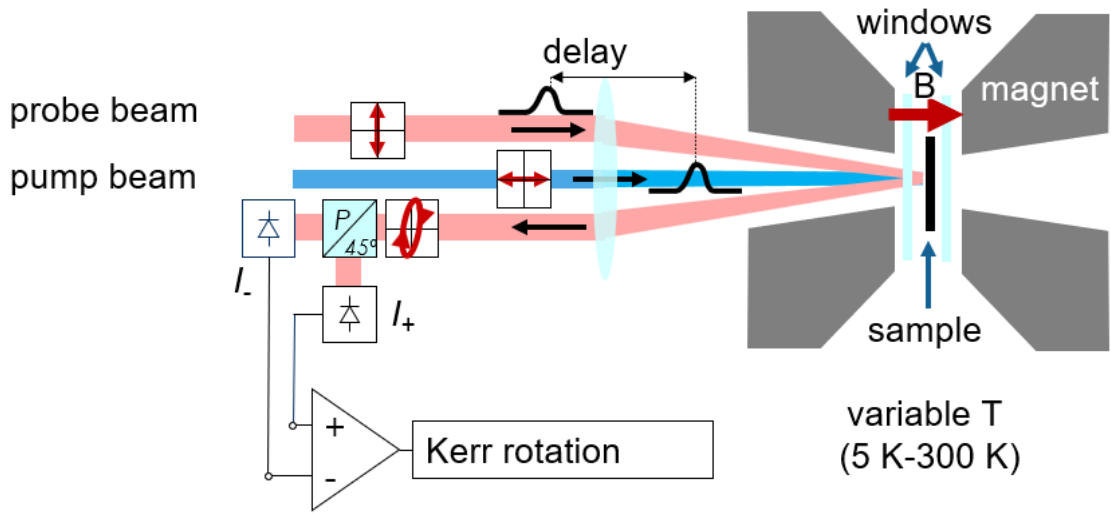

**Figure S21.** Schematics of the time-resolved magneto-optical Kerr effect measurement set up. A linearly-polarised pulsed pump beam, depicted in blue, is focused on a sample by a lens. To probe the OOP sample magnetization a delayed linearly-polarised pulsed probe beam, depicted in light red, is reflected from the sample. The magnetic field is pointing in OOP orientation with respect to sample. The polarisation of the reflected probe beam is then analysed by means of the polarising beam splitter and a pair of photodiode detectors. The four-quadrant squares with the red lines indicate the polarization of the beams at different positions along the path.

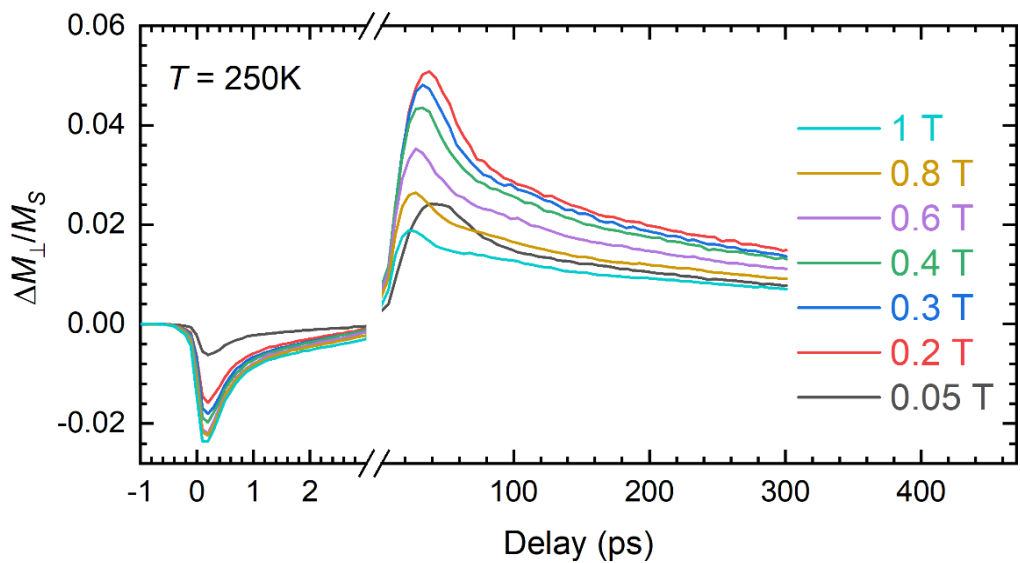

**Figure S22.** Magnetic-field dependent time-resolved dynamics at a pump fluence of $0.12 mJ/m^2$ where the magnetic field direction is pointing OOP with respect to sample as depicted in Figure S21. The demagnetisation (0-3ps) is followed by an increase of the OOP magnetisation, observed as the positive part of the magneto-optical Kerr angle transients (20-300ps) measured in different external fields at 250K in a Pt/Co(1.4nm)/H₂Pc system. The risetime is decreasing with increasing OOP magnetic field indicating precession dynamics.

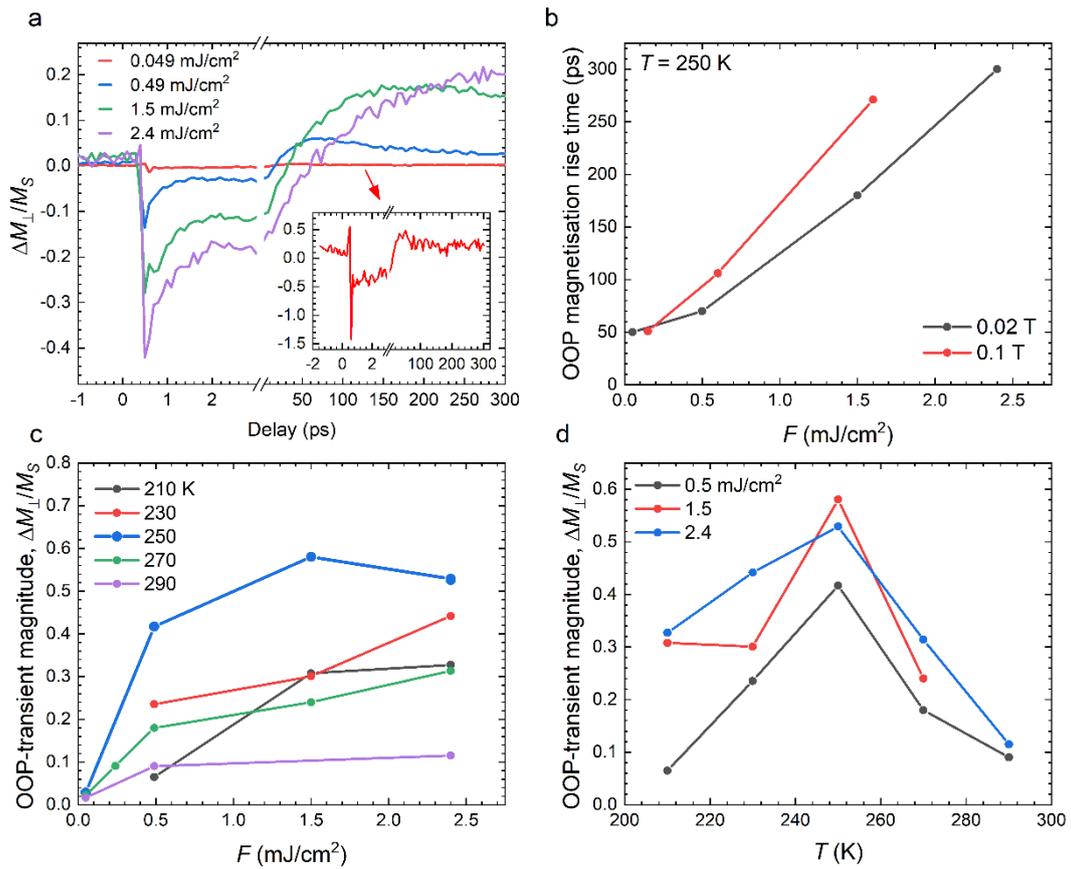

**Figure S23.** Pump-fluence and temperature dependence of the optical-pulse-induced OOP magnetisation transients. **a**, Ratio of onset OOP magnetisation to saturation magnetisation at different fluences at 250K in a weak OOP magnetic field of 0.02 T. **b**, The OOP-magnetisation rise time (the peak transient Kerr angle delay) as a function of the pump fluence at 250K at different OOP magnetic fields of 0.02 T and 0.1 T. **c**, The peak OOP magnetisation swing normalised to the static saturated magnetization ($M_S$) as a function of the pump fluence at different temperatures, showing saturation at higher fluences. **d**, The peak OOP magnetisation ($\Delta M_\perp/M_S$) as a function of temperature at different fluences, showing maximal OOP magnetisation swing to be at around 250K in the particular sample, irrespective of the fluence.

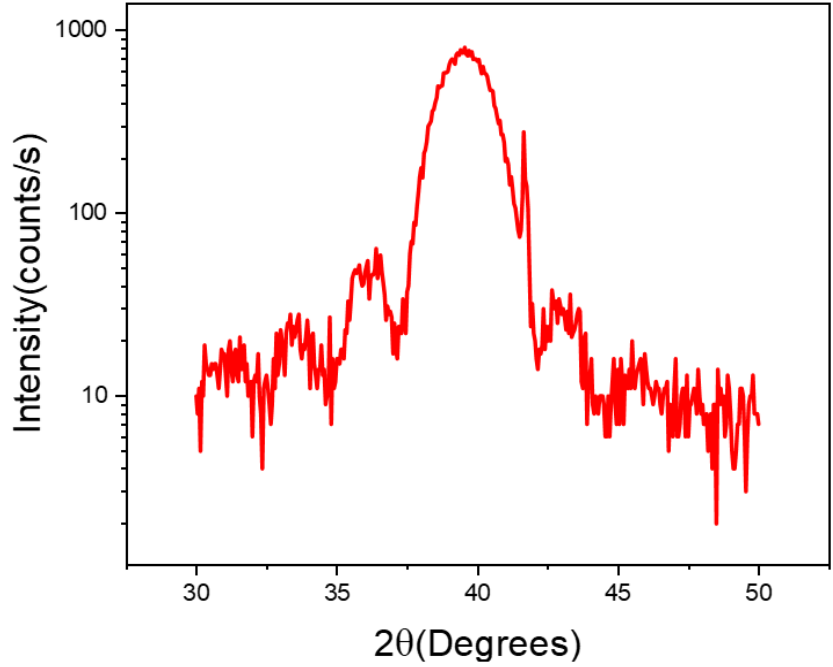

**Figure S24.** Pt(111) X-ray diffraction peak. Pendellosung fringes are visible showing the highly crystalline textured nature of the Pt seed layers.

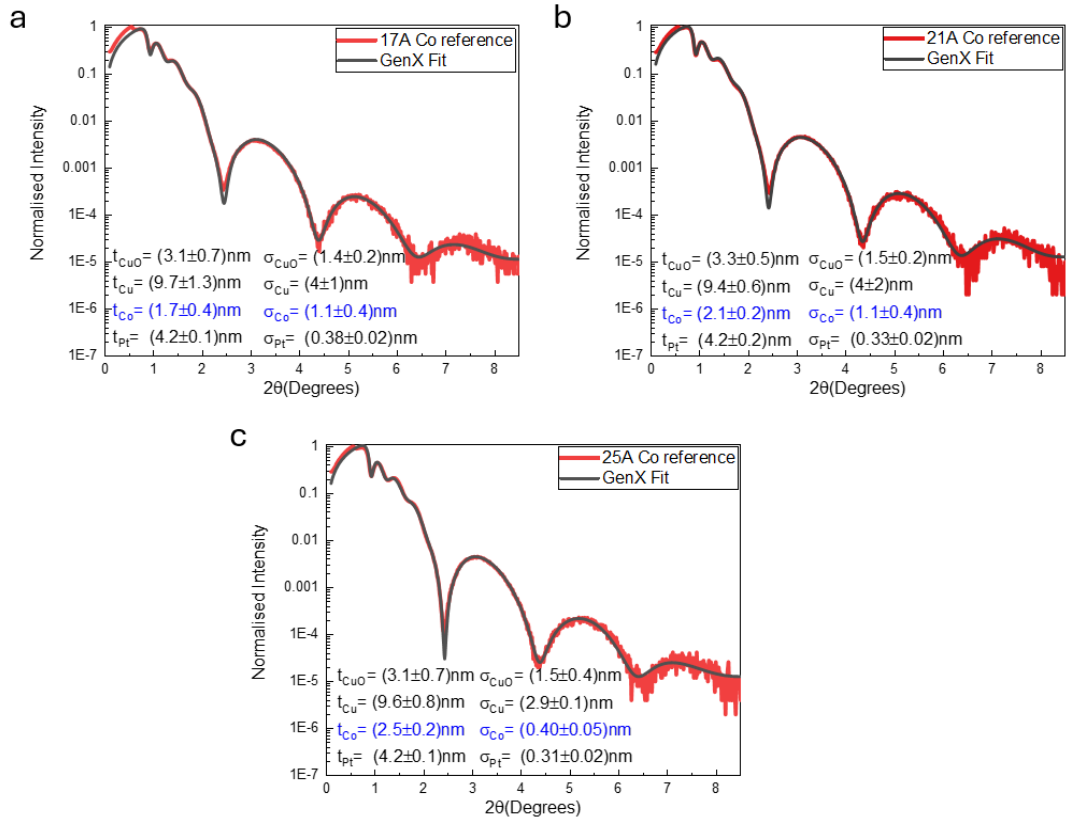

**Figure S25.** X-ray reflectivity characterisation. Reference Cu capped Pt/Co films of varying Co thickness **a**, 1.7nm, **b**, 2.1nm, and **c**, 2.5nm yielding fits in agreement with quartz crystal monitored thicknesses.

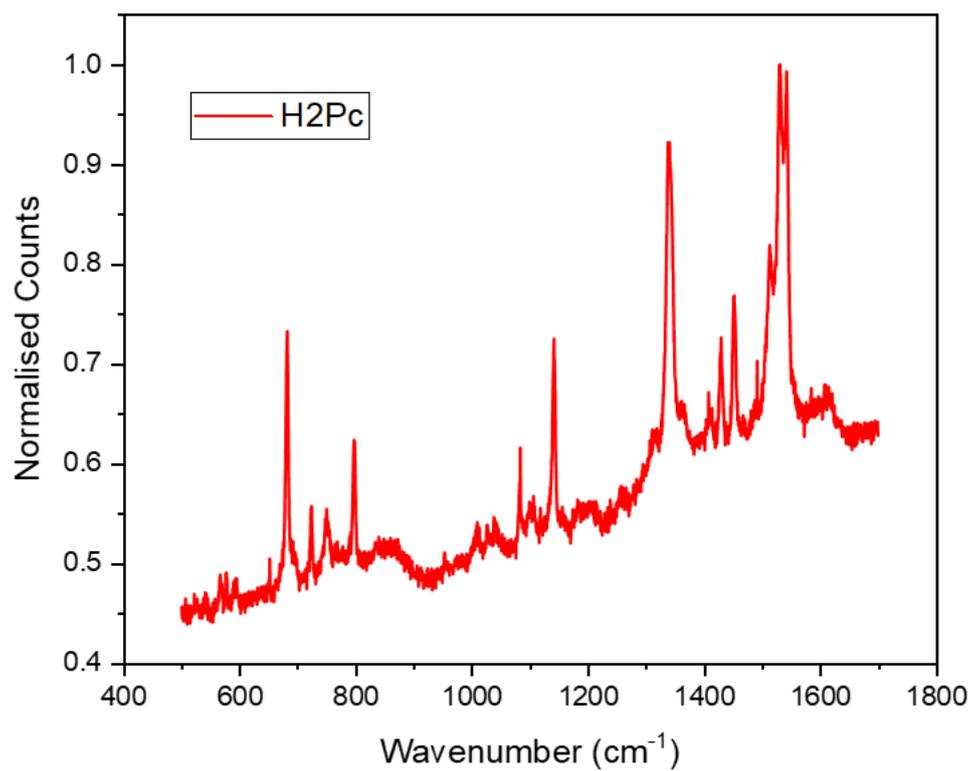

**Figure S26.** Raman spectroscopy of metallo-molecular structure. Characteristic Raman modes of Pc molecules evident on a capped Pt/Co/H₂Pc structure.